\pdfoutput=1
\documentclass[aps,twocolumn,superscriptaddress, prl]{revtex4-2}
\usepackage{amsmath}
\usepackage{amssymb}
\usepackage{graphicx}
\usepackage{epstopdf}
\usepackage[colorlinks=true]{hyperref}
\usepackage{physics}
\usepackage{mathrsfs}
\usepackage{comment}

\renewcommand\({\begin{equation}}	
\renewcommand\){\end{equation}}
\renewcommand\[{\begin{eqnarray}}	
\renewcommand\]{\end{eqnarray}}

\DeclareFontFamily{OT1}{pzc}{}
\DeclareFontShape{OT1}{pzc}{m}{it}{ <-> s*[1.26] pzcmi7t }{}
\DeclareMathAlphabet{\mathpzc}{OT1}{pzc}{m}{it}

\newcommand{\al}[1]{\begin{aligned}#1\end{aligned}}
\usepackage[dvipsnames]{xcolor}

\begin{document}

\title{Tunable Ultrafast Dynamics of Antiferromagnetic Vortices in Nanoscale  Dots}

\author{Ji Zou}
\affiliation{Department of Physics, University of Basel, Klingelbergstrasse 82, 4056 Basel, Switzerland}
\author{Even Thingstad}
\affiliation{Department of Physics, University of Basel, Klingelbergstrasse 82, 4056 Basel, Switzerland}
\author{Se Kwon Kim}
\affiliation{Department of Physics, Korea Advanced Institute of Science and Technology, Daejeon 34141, Republic of Korea}
\author{Jelena Klinovaja}
\affiliation{Department of Physics, University of Basel, Klingelbergstrasse 82, 4056 Basel, Switzerland}
\author{Daniel Loss}
\affiliation{Department of Physics, University of Basel, Klingelbergstrasse 82, 4056 Basel, Switzerland}

\begin{abstract}
Topological vortex textures in magnetic disks have garnered great attention due to their interesting physics and diverse applications. However, up to now, the vortex state has mainly been studied in microsize ferromagnetic disks, which have oscillation frequencies confined to the GHz range. Here, we propose an experimentally feasible \textit{ultrasmall} and \textit{ultrafast} vortex state in an antiferromagnetic nanodot surrounded by a heavy metal, which  is further harnessed to construct a highly tunable vortex network. We theoretically demonstrate that, interestingly, the interfacial Dzyaloshinskii-Moriya interaction (iDMI) induced by the heavy metal  at the boundary of the dot acts as an effective chemical potential for the  vortices in the interior. Mimicking the creation of a superfluid vortex  by rotation, we show that a magnetic vortex state can be stabilized by this iDMI. Subjecting the system to an electric current can trigger vortex oscillations via spin-transfer torque, which reside in the THz regime and can be further modulated by external magnetic fields. Furthermore, we show that coherent coupling between vortices in different nanodisks can be achieved via an antiferromagnetic link. Remarkably, this interaction depends on the vortex polarity and topological charge and is also exceptionally tunable through the vortex resonance frequency. This opens up the possibility for controllable interconnected networks of antiferromagnetic vortices.
 Our proposal therefore introduces a new avenue for developing  high-density memory, ultrafast logic devices, and THz signal generators, which are  ideal for  compact integration into microchips.
\end{abstract}

\date{\today}
\maketitle


\underline{\textit{Introduction.}}|A main theme in modern spintronics is investigating the potential of  nonlinear topological magnetic textures for innovative approaches in both classical and quantum information storage~\cite{parkin2008magnetic,TopologyinMagnetism,PhysRevLett.121.127701,liu2018binding,tang2021magnetic,zarzuela2020stability,cheng2019magnetic,psaroudaki2023skyrmion}, processing~\cite{christina_prl_2021,lan2015spin,Zou2023prr,jin2023nonlinear,xia2023universal,daniels2019topological,PhysRevB.97.064401,PhysRevB.95.144402,wang2018theory}, and transmission~\cite{jiprl2020,PhysRevLett.123.147203,Zangprl,Yqwinding,jivortex,zarzuela2023spin,yamane2019dynamics,dao2023topological}.  A prominent example that has attracted significant attention in the community
 is the magnetic vortex texture,  which is a topological spin configuration occurring in microscale ferromagnetic disks~\cite{shinjo2000magnetic,wachowiak2002direct}. It offers promising prospects for a broad spectrum of applications, including stable microwave signal generation~\cite{dussaux2010large,wintz2016magnetic,choi2007}, novel magnetic memory storage~\cite{yamada2007electrical,PhysRevLett.100.027203,pigeau2010frequency,bohlens2008current,hrkac2015magnetic}, logic devices~\cite{jung2012logic,shreya2022memory,PhysRevApplied.2.044001}, and reservoir and neuromorphic computing~\cite{torrejon2017neuromorphic,yun2022magnetic,shreya2023granular}. Much progress has been made  in the control and manipulation of the vortex state, as well as in exploring these diverse applications~\cite{yamada2007electrical,van2006magnetic,PhysRevLett.101.197204,liu2007current,fu2018optical,yamada2010current,PhysRevB.80.100405,kammerer2011magnetic,PhysRevLett.98.117201,kim2017fast,shibata2006current,gao2023interplay,li2021third}.
 However, all these studies have been restricted to vortex states in magnetic disks that possess finite net magnetizations. This restriction arises because the only known mechanism that can stabilize vortex configurations in microsize (or sub-microsize) disks is based on the dipolar interaction that requires the finite magnetization, which in turn limits the vortex oscillation frequency to the GHz or sub-GHz regime~\cite{pribiag2007magnetic,dussaux2010large,yu2013resonant,PhysRevLett.105.037201,PhysRevLett.106.197203}. Additionally, the microscale footprint of these textures not only hinders high-density integration in microchips but also poses challenges for their application in emerging quantum  technologies due to the rapid loss of quantum coherence associated with such large textures. 
 
 \begin{figure}
\includegraphics[width=0.96\columnwidth]{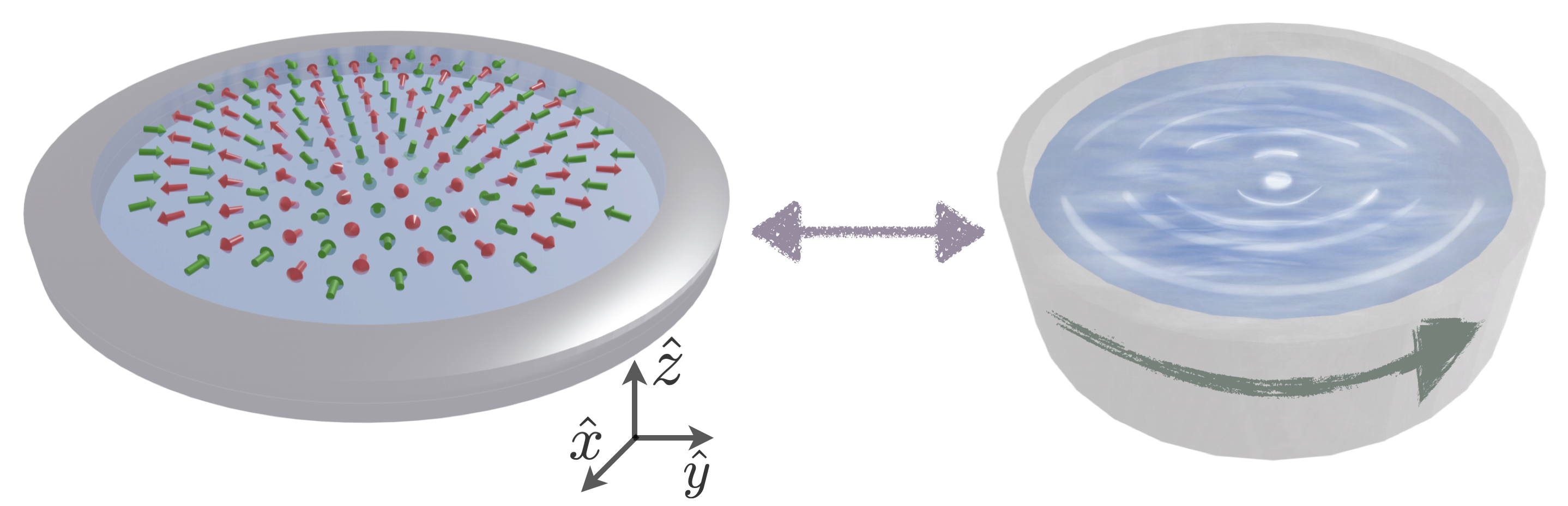}
\caption{A schematic of an antiferromagnetic disk (blue) surrounded by a heavy metal (grey) shown in the left panel. The iDMI at the boundary stabilizes vortex configurations in the interior. This is analogous to the formation of a vortex in a superfluid (blue) by rotation shown in the right panel.}
\label{fig1}
\end{figure}
 
In this work, we provide a different pathway to realize a vortex state by exploiting the interfacial Dzyaloshinskii-Moriya interaction (iDMI) in the system, which allows us to achieve a stable, \textit{ultrafast}, and \textit{ultrasmall} vortex configuration in magnetic disks even without net magnetization. We consider an experimentally feasible structure consisting of an antiferromagnetic disk surrounded by a heavy metal, as depicted in the left panel of Fig.~\ref{fig1}. In such a structure, the heavy metal provides an iDMI at the boundary of the disk due to the broken inversion symmetry  and the strong spin-orbit coupling in the heavy metal~\cite{RevModPhys.95.015003}. Interestingly, we demonstrate that this interaction acts as an effective chemical potential for vortices in the disk, akin to the rotation of superfluids that is known to induce vortex states. We show that a stable magnetic texture with finite vorticity, exhibiting vortex oscillation frequencies in the THz regime, is achievable in a disk with a diameter of ten nanometers, which would otherwise be impossible due to the dipolar interaction.
The vortex state responds to external magnetic fields, which break the time-reversal symmetry, lifting the degeneracy between the  left and right circular modes of vortex motion. We further show that electric currents can effectively trigger the ultrafast vortex oscillations exhibiting high quality factors, which thereby serve as stable generators of THz signals.

Exploiting magnetic vortices for unconventional computing and logic operations requires interaction between  disks in an interconnected magnetic vortex network~\cite{jung2012logic}. We demonstrate that this is achievable via coherent antiferromagnetic links. Importantly, we find that the coherent interaction, mediated by virtual magnons in the link, depends on both the polarity and topological charge of the two vortices. Furthermore, we demonstrate that this interaction is highly tunable through an applied magnetic field controlling the vortex resonance frequency.
Besides offering a nanoscale THz signal generator, our findings also provide opportunities for the development of new types of vortex-based memory and ultra-fast logic devices that combine storage and computation, allowing for high-density integration in microchips due to their nanoscale size.

\underline{\textit{General idea.}}|We first discuss how a vortex can emerge as the natural ground state in a magnetic system with  vanishing net magnetization.  To this end, we consider a thin, easy-plane antiferromagnetic dot of arbitrary shape, surrounded by a heavy metal. The left panel in Fig.~\ref{fig1} shows a circular geometry as an example. The dynamics of the antiferromagnet is described by  a smooth  N{\'e}el vector field $\vb n(x,y,t)$, which lives on the sphere $S^2$ for temperatures below the N{\'e}el temperature.  We introduce another smooth vector field,
\( \mathcal{A}_i=\frac{1}{2\pi} \hat{z}\cdot (\vb n \times \partial_i \vb n),  \)
which physically  describes the \textit{spin winding} of the  N{\'e}el vector field in the $xy$-plane (with $i\in\{x, y\}$)~\cite{Yqwinding}. In the limit of strong easy-plane anisotropy such that the N{\'e}el vector is confined to the $xy$-plane, this field $\vb*{\mathcal{A}}$ simplifies to $\mathcal{A}_i=(1/2\pi)\partial_i \phi$, with $\phi$ being the polar in-plane angle of $\vb n$ relative to $\hat{x}$. The \textit{vorticity density}  associated with the  N{\'e}el vector field then can be constructed as
\( \rho_v= \partial_x \mathcal{A}_y-\partial_y\mathcal{A}_x,  \)
which is a smooth function for smooth magnetic textures~\cite{quantumvortex}. We remark that, in the $xy$-plane limit,  it reduces to $\rho_v= (1/2\pi) \epsilon^{ij}\partial_i\partial_j \phi$  with the Levi-Civita tensor convention $\epsilon^{xy}=1$. The vorticity density is therefore zero everywhere except at isolated points with singular vortex cores.

The potential energy governing the energetics of the  N{\'e}el vector field comprises two parts: The first is the bulk energy $\mathcal{U}_0[\vb n]$ and the second is the iDMI arising from the strong spin-orbit coupling in the heavy metal and the absence of inversion symmetry at the interface~\cite{RevModPhys.95.015003,PhysRevB.102.224414,PhysRevLett.120.197202,volkov2018mesoscale}. These two parts together give rise to the total energy potential, which can be written into the following form:
\(  \mathcal{U}[\vb n]=\mathcal{U}_0[\vb n]-2\pi \mathcal{D} \oint_{\partial \Omega} d\vb*{\ell} \cdot \vb*{\mathcal{A}},   \)
where $\mathcal{D}$ is the iDMI strength and the curve integral is over the boundary $\partial\Omega$ of the magnetic dot. Note that this potential can be rewritten in the more suggestive form $\mathcal{U}=\mathcal{U}_0-2\pi\mathcal{D} \mathcal{N}$ by applying  Stokes' theorem, where $\mathcal{N}=\int_{\Omega} dxdy \, \rho_v$ is the total vortex number within the magnetic dot. Thus, the iDMI acts as an \textit{effective chemical potential}  for vortices in the antiferromagnetic dot|increasing $\mathcal{D}$ leads to a ground state with finite vorticity.

We remark that this bears close analogy with a rotating superfluid as shown in Fig.~\ref{fig1}. In the rotating frame, the superfluid acquires an additional potential term $\sim -\hbar \omega_0 \mathcal{N}$ originating from the Coriolis force~\cite{pethick2008bose}, with rotational frequency $\omega_0$ and total vorticity $\mathcal{N}$. It is known that the ground state of a superfluid transitions from vortex-free to one with a finite number of vortices above a critical rotational frequency~\cite{pethick2008bose}.
In our case, the iDMI plays the role of the rotational energy $\hbar \omega_0$.  Thus, we conclude that an antiferromagnetic state with finite vorticity forms when the iDMI strength exceeds a critical threshold, $\mathcal{D}>\mathcal{D}_c$. This  critical value can be approximately determined by balancing the iDMI and  the bulk energy $\mathcal{U}_0$ associated with a vortex configuration. While it is  a non-universal   parameter depending on the  energetic and geometric details of the magnet, we  expect $\mathcal{D}_c$ to be proportional to the magnetic exchange coupling which usually dominates the energy of nonlinear magnetic textures.

\underline{\textit{Model.}}|To illustrate the sketched mechanism, we consider a thin easy-plane antiferromagnetic circular dot for concreteness, as shown in Fig.~\ref{fig1}, with microscopic Hamiltonian $H=J\sum_{\langle i,j\rangle} \vb S_i\cdot \vb S_j+K \sum_i (\vb S_i\cdot \hat{z})^2$. Here $J>0$ stands for the antiferromagnetic exchange coupling between two nearest sites $\langle i, j\rangle$, and $K>0$ defines the hard $\hat{z}$-axis.  The  low-energy dynamics of the system is captured by the  N{\'e}el vector field  $\vb n$ with the effective Lagrangian $\mathcal{L}[\vb n]=\int dxdy\, \varrho\, \dot{\vb n}^2 /2 - \mathcal{U}_0[\vb n]$,  where $\varrho = \hbar t_z/(2Jza^3)$ is the effective inertia with lattice spacing $a$, the thickness in the $\hat{z}$-direction $t_z$, and  the coordination number $z$. The potential energy is 
\(  \mathcal{U}_0[\vb n] = \int\! dxdy\; \left[ \frac{\mathcal{J}}{2} (\nabla \vb n)^2 + \frac{\mathcal{K}}{2} (\vb n \cdot \hat{z})^2 \right]. \)
Here, the exchange stiffness 
$\mathcal{J}= JS^2t_z/a$ 
and the hard $\hat{z}$ -axis anisotropy $\mathcal{K}= KS^2/a^2$  (with spin length $S$) penalize states with  non-uniform spatial textures and states with $\vb n$ deviating from the $xy$-plane, respectively. 
Together, they define a   characteristic length scale $\lambda_0=\sqrt{\mathcal{J}/\mathcal{K}}$ for the size of vortices in the dot. We consider an antiferromagnetic disk with radius comparable to the vortex size, $R\approx \lambda_0$~\cite{afm_radius}. 
To leverage the circular symmetry of the system, we use polar coordinates  $(r, \varphi)$, and further adopt the convenient parametrization $\vb n(r,\varphi)=(\sin\theta\cos\phi, \sin\theta\sin\phi, \cos\theta)$ for the spin texture.

We now wish to examine whether the system   favors a ground state with finite vorticity. To this end, we employ the ansatz~\cite{guslienko2001field} $\theta=2\arctan (r/\lambda_0)$ for $r<\lambda_0$ (otherwise $\theta=\pi/2$) and $\phi=\mathcal{N}\varphi$, which gives a profile with positive polarity $p\equiv n_z(r=0)=+1$ and vortex number $\mathcal{N}$. This yields a total energy of $\mathcal{U}\propto \mathcal{J} \mathcal{N}^2-2\mathcal{D} \mathcal{N}$~\cite{afm_value}. Thus, in the presence of the sizable iDMI (comparable to $\mathcal{J}$), the system ground state changes from  a vortex-free configuration to one with finite vorticity. Considering a NiO antiferromagnetic dot with a thickness of $t_z=2\,\text{nm}$, the exchange stiffness is approximately  $\mathcal{J}\approx 10^{-21}\, \text{J}$~\cite{PhysRevLett.112.147204}. The interfacial DM interaction is about $\mathcal{D}\approx 4\times 10^{-21} \,\text{J}> \mathcal{J}$ for typical heavy metals such as W or Pt (assuming a penetration depth of $2\, \text{nm}$)~\cite{PhysRevB.102.224414}. With the hard $\hat{z}$-axis anisotropy being $\mathcal{K}\approx 4.5\times 10^{-5}\text{J}/\text{m}^2$~\cite{PhysRevB.95.104418}, the characteristic vortex size (or the radius of the magnetic dot) is evaluated to be around 5 nm. Therefore, an antiferromagnetic vortex state is experimentally feasible in a thin nanoscale magnetic dot.

One important parameter of the system is the characteristic frequency associated with the vortex oscillation in the dot. We consider a configuration $\vb n(\vb r, t)=\vb n_0(\vb r-\vb*{\mathcal{R}})$ with $\vb n_0$ being the previous ansatz with $\mathcal{N}=1$. Here, $\vb*{\mathcal{R}}(t)$ describes the time-dependent vortex position, and the associated mass $\mathcal{M}\equiv 2\pi \varrho$ follows from the Lagrangian for the N{\'e}el vector field  through $\int d\vb r\, \varrho \dot{\vb n}^2/2=\mathcal{M}\dot{\vb*{\mathcal{R}}}^2/2$.
 The response of the vortex  to a small  displacement can be determined by varying the potential energy with respect to $\vb*{\mathcal{R}}$. The variation of the bulk potential leads to an inverted harmonic potential $\delta \mathcal{U}_0\approx -\pi \mathcal{J} (\vb*{\mathcal{R}}/\lambda_0)^2$, indicating interactions within the dot disfavor the vortex, pushing it out of the system. Remarkably, the variation of the iDMI generates a harmonic trap $\pi \mathcal{D}(\vb*{\mathcal{R}}/\lambda_0)^2$  for the vortex, effectively retaining it within the system. When the  iDMI falls below the critical value $\mathcal{D}_c\equiv \mathcal{J}$, the  vortex at the origin would escape from the dot even under a small perturbation.   The total vortex potential energy  is given by
\( \delta \mathcal{U}\approx\frac{1}{2}\mathcal{M} \Omega_0^2 \vb*{\mathcal{R}}^2, \;\;\text{with}\;\; \Omega_0^2= \frac{2\pi (\mathcal{D}-\mathcal{D}_c)}{\mathcal{M}\lambda_0^2}.  \)
This vortex oscillation frequency is estimated to be about $\Omega_0=2\pi \times 0.26\, \text{THz}$ for NiO, which is several orders of magnitude larger than the existing ferromagnetic vortex oscillation frequency.

\begin{figure}
\includegraphics[width=\columnwidth]{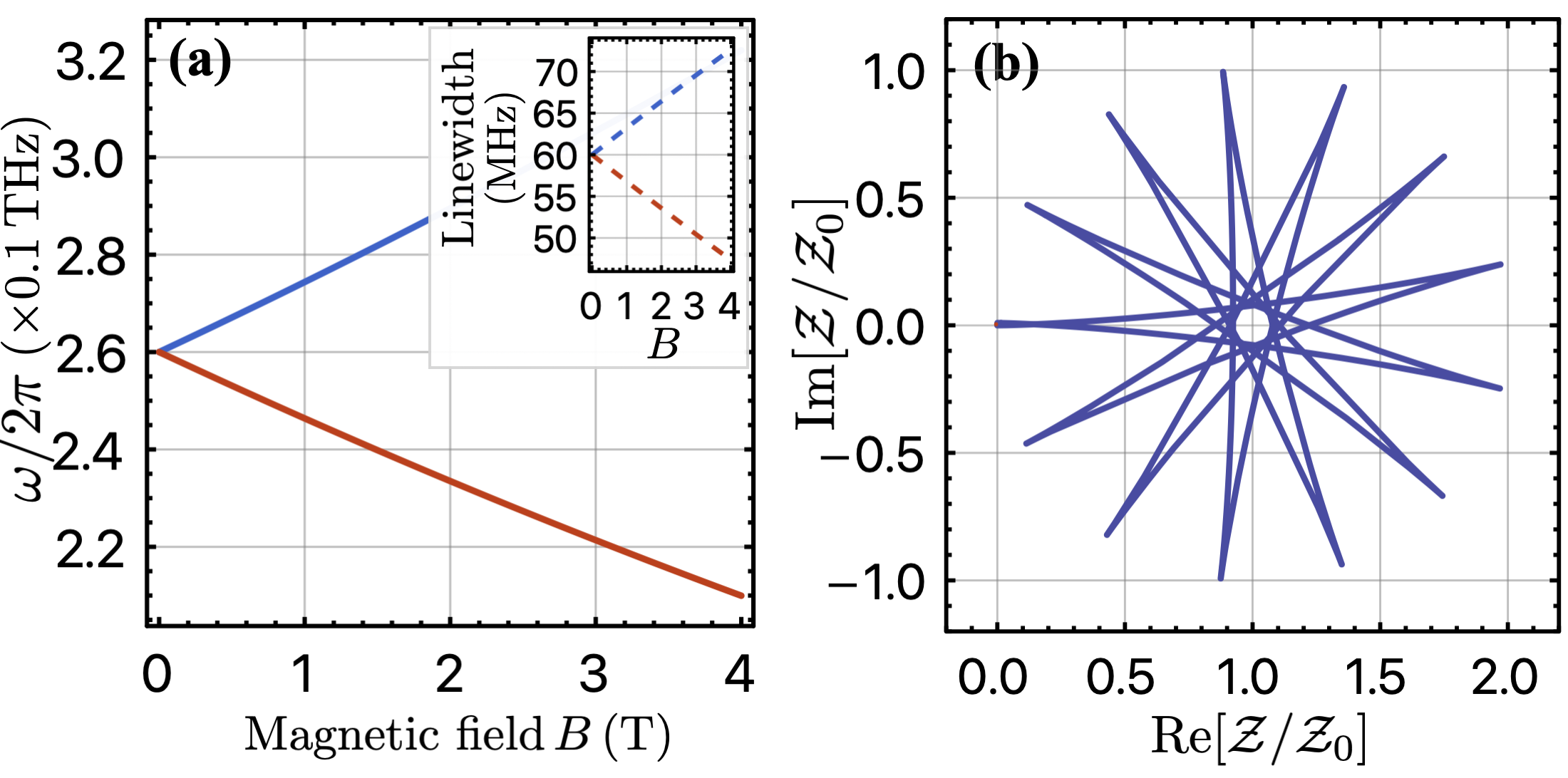}
\caption{Vortex dynamics. (a) Dependence of left and right circular vortex oscillation frequencies on the magnetic field. With increasing  field, $\omega_L$ (red line) decreases while $\omega_R$ (blue line) increases. The inset depicts the linewidth of $\omega_R$ (blue dashed line) and $\omega_L$ (red dashed line) as a function of $B$ field,  exhibiting asymmetry at finite magnetic fields.  (b) Trajectory of the vortex. Here we set $\gamma=0$ and $\omega_c=20\,\text{GHz}$. The current is assumed to be along the $y$ direction. }
\label{fig2}
\end{figure}

\underline{\textit{Magnetic field and spin-transfer torque.}}|Magnetic fields and electric currents stand out as the most effective tools for manipulating magnetic states in modern spintronics~\cite{bader2010spintronics,RevModPhys.90.015005}. Here we wish to study the response of the antiferromagnetic vortex state to these external fields. The kinetic term of the Lagrangian for the N{\'e}el vector field   reads $\int d^2\vb r\, \varrho (\partial_t{\vb n}-\vb h \times \vb n)^2/2$ in the presence of the Zeeman coupling $-\hbar \sum_i\vb h \cdot \vb S_i$, where $\vb h\equiv \gamma_e \vb B$ with magnetic field $\vb B$ and  gyromagnetic ratio $\gamma_e$. We assume the applied field to be perpendicular to the magnetic dot, $\vb h=h\hat{z}$. Introducing a gauge field $\mathpzc{A}_i=\varrho h \int d\vb r\, \hat{z}\cdot ( \vb n \times \vb \partial_{i} \vb n )$, the cross term linear in $\vb h$ takes the form $\mathpzc{A}_i \dot{\mathcal{R}}_i$. It generates a Magnus force $\dot{\vb*{\mathcal{R}}}\times \mathpzc{B}\hat{z}$ acting on the vortex through the effective field
\( \mathpzc{B}=\epsilon_{ij} \partial_{\mathcal{R}_i}\mathpzc{A}_j= -\mathcal{M} h\mathcal{N}. \)
 When an electric current flows through the metallic antiferromagnetic dot, the s-d coupling between the local magnetic moments and the electron spins polarizes the electrons to follow the orientation of spins on individual sublattices, leading to the adiabatic spin-transfer torque on each sublattice~\cite{PhysRevLett.113.057601,PhysRevLett.106.107206}. This effect can be captured by replacing the time derivative with a convective derivative $D_t=\partial_t+\vb u\cdot \nabla$ in the kinetic term~\cite{PhysRevB.83.054428}, where $\vb u$ is the drift velocity of the electrons in the current.
  This leads to a nontrivial  term $\varrho \int d\vb r\, \vb h\cdot[ (\vb u\cdot \nabla) \vb n \times \vb n]$, which gives  rise to  a Magnus force acting on the vortex  transverse to the applied current $-\vb u\times \mathpzc{B}\hat{z}$~\cite{ji_2021_neel,sayakgauge}. It is   analogous to the Magnus force exerted on a vortex line in superconductors when subjected to an electric current~\cite{RevModPhys.36.45}.  Now, the equation of motion for the vortex in the antiferromagnetic disk is
\( \mathcal{M}\ddot{\vb*{\mathcal{R}}}  + 2\mathcal{M} \gamma \dot{\vb*{\mathcal{R}}} +(\vb u- \dot{\vb*{\mathcal{R}} } )\times \mathpzc{B} \hat{z} +\nabla_{\vb*{\mathcal{R}}} \delta \mathcal{U}=0,\)
where the second term accounts for the friction  captured by the Rayleigh dissipation function $(\hbar\alpha s/2)\int d^2\vb r \,\dot{\vb n}^2$ with the Gilbert damping $\alpha$ and spin density $s$. The friction coefficient  $\gamma\equiv \pi \hbar \alpha s/\mathcal{M}$ translates into the linewidth of the vortex oscillation modes, which is about $\gamma\approx 2\pi \times 60 \, \text{MHz}$ when $\alpha\sim 10^{-4}$ and $s\sim 1/a^2$. The vortex oscillator is therefore  characterized by a high quality factor $\Omega_0/\gamma\gtrsim 4\times 10^3$ and  well-suited for stable and coherent THz signal generation.

The response function of the system $\hat{\chi}(\omega)$ is defined through $\vb*{\mathcal{R}}^T(\omega)=\hat{\chi}(\omega)\vb f^T(\omega)/\mathcal{M}$, where  $\vb f$ is an external force. From the equation of motion,  it is given by 
\(  \hat{ \chi}^{-1}(\omega)= \chi_0^{-1}(\omega)  + 2\omega_c \omega \hat{\sigma}_y ,   \)
where $\omega_c\equiv -\mathpzc{B}/(2\mathcal{M})=h\mathcal{N}/2$ is the effective cyclotron frequency and $\chi_0(\omega)\equiv-(\omega^2+2i\gamma\omega-\Omega_0^2)^{-1}$ is the bare response function in the absence of an  applied magnetic field. 
The Pauli matrix $\hat{\sigma}_y$ originates from  the breaking of time-reversal symmetry. The poles of $\hat{\chi}$ yields the characteristic frequencies associated with the  right and left circular vortex-oscillation modes,
$\omega_{R/L} \approx \Omega \pm \omega_c $ with $\Omega =\sqrt{\Omega_0^2+\omega_c^2}$ and linewidth $ \gamma (1\pm \omega_c/\Omega )$.  Note  that the applied magnetic field lifts the degeneracy of the two circular modes, while also producing linewidth 
asymmetry.
 As depicted in Fig.~\ref{fig2}(a),  the frequency of the left (right) circular mode is pushed downward (upward) and attains  a narrower (broader) linewidth when the field is oriented along the positive $\hat{z}$-axis. 
 We point out that, importantly,  these modes can be excited by an applied in-plane dc current.  Consider the scenario where a vortex, initially stationary at the center of the disk, is subjected to a current characterized by an electron drift velocity $\vb u$. In this case, the current-induced vortex motion is  $\mathcal{Z}(t) = \mathcal{Z}_0 [1 - (\omega_R e^{-i\omega_L t} + \omega_L e^{i\omega_R t}) / 2\Omega]$, where $\mathcal{Z} \equiv \mathcal{R}_x + i\mathcal{R}_y$ represents the vortex position. 
  The finite linewidth ultimately brings the vortex to its stationary position $\mathcal{Z}_0= 2\omega_c (u_y-iu_x)/\Omega_0^2$, where the restoring force is balanced with the Magnus force exerted by the current. As an example, we depict the vortex trajectory in real space with vanishing damping in Fig.~\ref{fig2}(b) for an applied current in $y$ direction.  The final stationary position is on the $x$ axis with finite linewidth~\cite{afm_sm}.

\begin{figure}
\includegraphics[width=0.98\columnwidth]{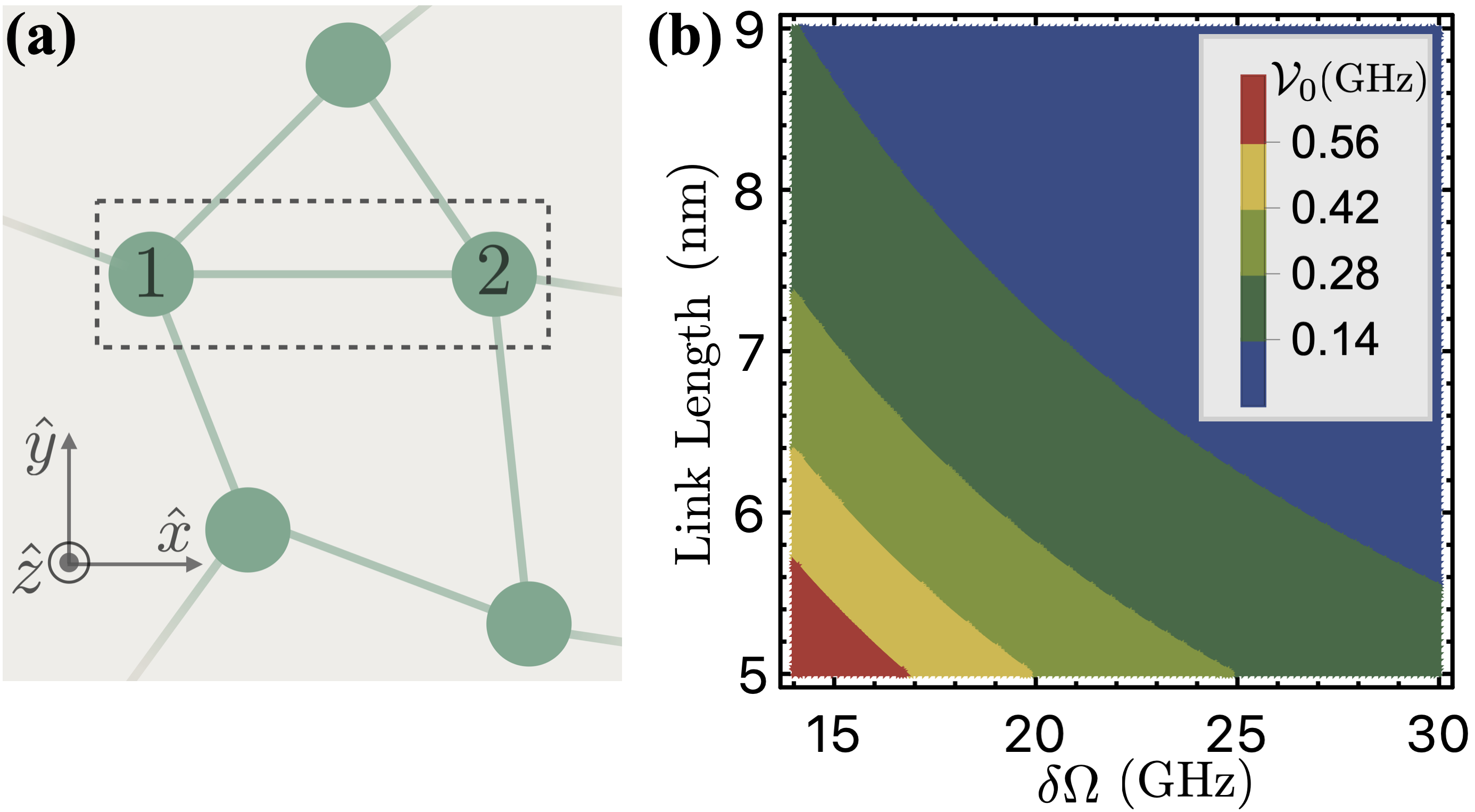}
\caption{(a) A patch of an interconnected vortex network. Disks, represented by green dots, are connected by antiferromagnetic links. (b) The coupling strength between vortices as a function of the link length and the frequency difference $\delta\Omega\equiv \Delta-\Omega_0.$ In the plot, we set the antiferromagnetic coupling in the link and the easy-axis anisotropy to be $J_A/\hbar\approx 2\pi \times 210\, \text{GHz}$ and $K_A/\hbar\approx 2\pi \times 40\, \text{GHz}$, respectively. We also set the coupling between the magnetic disk and the link to be $\mathcal{J}_A/\hbar\approx 2\pi \times 20 \,\text{GHz}$.}
\label{fig3}
\end{figure}

\underline{\textit{Tunable  vortex interactions.}}|A magnetic vortex network requires a coherent interaction between different vortex disks. Here we illustrate that such a coupling can be mediated via an antiferromagnetic link with gapped magnonic excitations, as sketched in Fig.~\ref{fig3}(a). To this end, we consider a microscopic Hamiltonian $H_{\text{link}}=J_A\sum_{\langle i,j\rangle} \vb S_i\cdot \vb S_j -K_A \sum_i (\vb S_i\cdot \hat{x})^2$ describing the quasi-one-dimensional antiferromagnetic coupler. The coefficients $J_A$ and $K_A$ are both positive, standing for the antiferromagnetic coupling and the easy-axis anisotropy along the link (defined as $\hat{x}$-axis), respectively. The interaction between vortices is primarily mediated by magnon excitations with small momentum $k$, and the corresponding spectrum is $\omega_k=A k^2 +\Delta$~\cite{afm_sm}.
 Here, $A\equiv J_ASa^2/(\hbar E_g)$ is the exchange stiffness and $\Delta\equiv 2S E_gJ_A/\hbar$ is the excitation gap, where $E_g=\sqrt{(1+K_A/J_A)^2-1}$. We consider two magnetic dots coupled to the left and right ends of the link through exchange coupling: \(H^\prime/\mathcal{J}_A=  \vb n_1 ({R}\hat{x}) \cdot \vb n_{\text{link}} (0) + \vb n_2 (-{R}\hat{x}) \cdot \vb n_{\text{link}} (L), \)
where $\mathcal{J}_A$ is the coupling strength and $\vb n_1 ({R}\hat{x})$, $\vb n_2 (-{R}\hat{x}) $ denote the N{\'e}el vectors at the right side of the first disk and the left side of the second disk, respectively; $\vb n_{\text{link}} (0)$, $\vb n_{\text{link}} (L) $ stand for the N{\'e}el vector field of the link (with total length  $L$) evaluated at two ends. We again adopt the collective coordinates for vortices, which yields $n_x(\pm\mathcal{R}\hat{x})\approx\pm 1-\mathcal{X}, n_y(\pm\mathcal{R}\hat{x})\approx- \mathcal{N} \mathcal{Y}, n_z(\pm\mathcal{R}\hat{x})\approx\pm p \mathcal{X}$~\cite{afm_sm} with dimensionless displacement $(\mathcal{X}, \mathcal{Y})$  in  units of the magnetic exchange length $\lambda_0$ to linear order in $\mathcal{X}$ and $\mathcal{Y}$. Here, $p, \mathcal{N}=\pm 1$ are the  polarity and   the charge of the vortex, respectively.

We focus on the scenario where the vortex oscillation frequency falls below the band gap $\Delta$, ensuring that the vortex dynamics do not excite magnons in the link.  Otherwise, the link could result in extra damping and dissipative coupling within the vortices~\cite{Daniel2013prx,hetenyi2022long,zou2024prl,zou2023spatially}. To derive the coherent vortex-vortex interaction mediated by the virtual magnons, we trace out the the degrees of freedom of the link by applying the Schrieffer-Wolff transformation~\cite{bravyi2011schrieffer}, which yields
\( \mathcal{V}\approx \mathcal{V}_0 \left[  -p_1 p_2 \mathcal{X}_1\mathcal{X}_2 + \mathcal{N}_1\mathcal{N}_2 \mathcal{Y}_1\mathcal{Y}_2   \right].  \)
Note that, importantly,  both polarity and  topological number can change the sign of the interaction. We have focused on the case without external magnetic field for simplicity.   The detailed derivation and general discussion are provided in SM~\cite{afm_sm}.   The coupling strength $\mathcal{V}_0=\mathcal{J}_A^2\chi_\perp(\Omega_0, L)/(2\hbar)$ is determined by the  transversal susceptibility  of the antiferromagnetic link evaluated to be
\(  \chi_\perp (\Omega_0,L)\approx  \frac{\alpha_0 e^{-k_F L}   }{A(k_F /a)},\;\;\text{with}\;\; k_F=\sqrt{\frac{\Delta-\Omega_0}{A}}, \)
where $\alpha_0$ is a dimensionless factor of order one and  $1/k_F$ defines a  characteristic length that increases as the vortex oscillation frequency approaches the magnon band edge. As we show in Fig.~\ref{fig3}(b), the coupling $\mathcal{V}_0$ decays as $\exp(-k_FL)$ when the link length increases. On the other hand, for given $L$, the coupling scales nonlinearly with the vortex oscillation frequency: $\mathcal{V}_0\propto \exp[-(L/\sqrt{A})\sqrt{\delta \Omega}]/\sqrt{\delta\Omega}$ with $\delta\Omega=\Delta-\Omega_0$. This offers a convenient way to tune the coupling between vortices by adjusting the vortex resonance frequency, for instance, via  a magnetic field. We finally point out that, while reducing $\delta \Omega$ can enhance the coupling, we still require $\delta \Omega$ to be comparable or  larger than $\mathcal{J}_A$ to validate the perturbation theory  employed in our discussion~\cite{bravyi2011schrieffer}.

\underline{\textit{Summary and outlook.}}|We explored the feasibility of sustaining a stable vortex state in an antiferromagnetic nanodot encircled by a heavy metal. We reveal that in the presence of iDMI, a vortex state can be stabilized, akin to the formation of a vortex by rotation in a superfluid.  The vortex state serves as a stable generator of coherent THz signals, which can be further modulated with magnetic fields and effectively excited by electric currents, heralding a range of innovative applications. Furthermore, by linking two nanodots through an antiferromagnetic coupler, virtual magnons enable tunable interactions between vortices in the dots.

Building upon our study, one could envision a range of interesting future research directions. Firstly, in addition to investigating the realization of non-volatile logic operations and unconventional computation within a vortex network,  one could utilize nanodisks as meta-atoms to construct metamaterials, offering a fertile  ground for exploring various phenomena~\cite{li2021topological,PhysRevLett.119.077204,li2019higher,PhysRevB.98.180407,PhysRevB.103.054438}. Secondly, while our focus has been on the coherent coupling between vortices, future studies can  explore the dissipative coupling between vortices, which naturally occurs when the vortex resonance frequency falls within the magnon band of the coupler. This leads to  a dissipatively interconnected vortex network, providing an ideal platform for examining the utilization of dissipation or for investigating various non-Hermitian magnonic phases, which have recently garnered considerable interest in the community~\cite{yu2023non,hurst2022non,yu2019prediction,zou2022prb,ashida2020non,nakata2024magnonic}. Lastly, the nanoscale size of the vortex state not only facilitates high-density integration on microchips but also opens up possibilities for harnessing its quantum properties~\cite{YUAN20221}. For example, there is potential for using the vortex state as a qubit or integrating it into existing quantum computation platforms.

\begin{acknowledgments}
\textit{Acknowledgments.} This work was supported by the Georg H. Endress Foundation and by the Swiss National Science Foundation, NCCR SPIN (grant number 51NF40-180604). S.K.K. was supported by the Brain Pool Plus Program through the National Research Foundation of Korea funded by the Ministry of Science and ICT (2020H1D3A2A03099291).
\end{acknowledgments}

%

\onecolumngrid
\clearpage
\setcounter{equation}{0}
\renewcommand{\theequation}{S\arabic{equation}}
\renewcommand{\thefigure}{S\arabic{figure}}
\appendix

{\centering
    \large{\textbf{{Supplemental Material for\\ ``Tunable Ultrafast Dynamics of Antiferromagnetic Vortices in Nanoscale  Dots"}}}
\par}

\bigskip

\author{Ji Zou}
\affiliation{Department of Physics, University of Basel, Klingelbergstrasse 82, 4056 Basel, Switzerland}
\author{Even Thingstad}
\affiliation{Department of Physics, University of Basel, Klingelbergstrasse 82, 4056 Basel, Switzerland}
\author{Se Kwon Kim}
\affiliation{Department of Physics, Korea Advanced Institute of Science and Technology, Daejeon 34141, Republic of Korea}
\author{Jelena Klinovaja}
\affiliation{Department of Physics, University of Basel, Klingelbergstrasse 82, 4056 Basel, Switzerland}
\author{Daniel Loss}
\affiliation{Department of Physics, University of Basel, Klingelbergstrasse 82, 4056 Basel, Switzerland}

\maketitle

\subsection*{(i) Derivation of vortex-vortex interaction}
Here we provide detailed derivations and  discussions on the coherent interactions between vortices connected via an antiferromagnetic coupler. Let us first focus on the Hamiltonian describing the exchange coupling between the vortex disk and the link: 
\(H^\prime/\mathcal{J}_A=  \vb n_1 ({R}\hat{x}) \cdot \vb n_{\text{link}} (0) + \vb n_2 (-{R}\hat{x}) \cdot \vb n_{\text{link}} (L), \label{s1} \)
where $\mathcal{J}_A$ is the coupling strength. Here $\vb n_1 ({R}\hat{x})$ and  $\vb n_2 (-{R}\hat{x})$ stands for the N{\'e}el vector fields of the left and right magnetic disks evaluated at $\vb r=R\hat{x}$ and $\vb r=-{R}\hat{x}$ (with the origins at the center of the disks).  We adopt the collective coordinates $\vb n(\vb r, t)=\vb n_0(\vb r-\vb*{\mathcal{R}}(t))$, with $\vb n_0(\vb r)$ being the ansatz that we assume in the main text. It can be written as:
\(  n_x(\vb r)=\frac{2x/\lambda_0}{ (x/\lambda_0)^2+(y/\lambda_0)^2+1}, \;\;\;n_y(\vb r)=\frac{2y/\lambda_0}{ (x/\lambda_0)^2+(y/\lambda_0)^2+1},  \;\;\; n_z(\vb r)= \frac{2}{(x/\lambda_0)^2+(y/\lambda_0)^2+1}-1,  \)
in the Cartesian coordinate with $\lambda_0$ being the characteristic length. We recall that we consider a disk with a radius comparable to this length $R\sim \lambda_0$. Then we have:
\( \al{  n_x ({R}\hat{x}) &=  \frac{2(1-\mathcal{X})}{(1-\mathcal{X})^2 +\mathcal{Y}^2+1} \approx 1-\mathcal{X},  \\
          n_y ({R}\hat{x})  &= \frac{2(-\mathcal{Y})}{(1-\mathcal{X})^2 +\mathcal{Y}^2+1} \approx -\mathcal{Y},  \\
            n_z ({R}\hat{x})  &= \frac{2}{(1-\mathcal{X})^2 +\mathcal{Y}^2+1}  -1 \approx +\mathcal{X},   }\)
            where we have introduced the dimensionless coordinate $\vb*{\mathcal{R}}/\lambda_0\equiv (\mathcal{X}, \mathcal{Y})$ in units of $\lambda_0$. Similarly one can find that \( n_x (-{R}\hat{x}) \approx -1-\mathcal{X}, \;\;\; n_y (-{R}\hat{x}) \approx -\mathcal{Y}, \;\;\; n_z ({R}\hat{x}) \approx -\mathcal{X}. \)
            We remark that we have used the ansatz with positive polarity $p=+1$ and positive vortex number $\mathcal{N}=+1$. When the polarity flips, we have $n_z\rightarrow -n_z$, and when $\mathcal{N}=-1$, it results in $n_y\rightarrow -n_y$. Thus, the Hamiltonian~\eqref{s1} can be written as
            \(  \al{ H^\prime/\mathcal{J}_A&= (1-\mathcal{X}_1)n_x(0)+(-1-\mathcal{X}_2) n_x(L)-\mathcal{Y}_1n_y(0) -\mathcal{Y}_2n_y(L) +\mathcal{X}_1n_z(0) -\mathcal{X}_2n_z(L),  \\
              &=  (1-\mathcal{X}_1)n_x(0)+(-1-\mathcal{X}_2) n_x(L) + \frac{\mathcal{X}_1-i\mathcal{Y}_1}{2i}n^+(0) - \frac{\mathcal{X}_1+i\mathcal{Y}_1}{2i}n^-(0)  \\
                & \;\;\; \;\;\; \;\;\; +   \frac{\mathcal{X}_2-i\mathcal{Y}_2}{2i}n^-(L)  - \frac{\mathcal{X}_2+i\mathcal{Y}_2}{2i}n^+(L)    } \) 
            where we have dropped the sub-index `link' for notational convenience and introduced $n^+\equiv n_y+in_z$, $n^-\equiv n_y-in_z$.  To proceed, we  quantize the vortex motion in the terms of the operators for left and right circular modes $d_L$, $d_L^\dagger$ and $d_R$, $d_R^\dagger$.  Then we have
            \(  \mathcal{X}+i\mathcal{Y}=\sqrt{2} \lambda (d_R+d_L^\dagger),     \)
            where $\lambda=\sqrt{\hbar/(2\mathcal{M}\Omega)}/\lambda_0$ is the dimensionless harmonic length of the confining potential of the vortex (with $\Omega=\sqrt{\Omega_0^2+\omega_c^2}$). 
            
            We now focus on the  microscopic Hamiltonian $H_{\text{link}}=J_A\sum_{\langle i,j\rangle} \vb S_i\cdot \vb S_j -K_A \sum_i (\vb S_i\cdot \hat{x})^2$ of the quasi-one-dimensional (two-sublattice) antiferromagnetic coupler. Here, $J_A$ and $K_A$ are both positive, standing for the antiferromagnetic coupling and the easy $\hat{x}$ anisotropy along the link, respectively. We introduce the Holstein-Primakoff transformation and keep up to the lowest order: $S_A^+\approx\sqrt{2S}a$ and $S_A^x=S-a^\dagger a$ for site A and $S_B^+\approx\sqrt{2S} b^\dagger$ and $S_A^x=-S+a^\dagger a$ for site B with $S^{\pm}\equiv S_y\pm iS_z$. Then this Hamiltonian can be written as
            \( H_{\text{link}} = 2SJ_A\sum_{k} \left[ (K_A/J_A+1)(a_k^\dagger a_k + b_k^\dagger b_k) +\gamma_k(a_kb_k+a_k^\dagger b_k^\dagger)    \right],  \)
            in the momentum space with $\gamma_k=(e^{ik}+e^{-ik})/2=\cos k$ (we use dimensionless $k$ by setting the spacing $a=1$ for simplicity). It can be diagonalized by applying the Bogoliubov transformation  
            \(  a_k=u_k \alpha_k +v_k \beta_k^\dagger, \;\;\;\;\;\; b_k = u_k \beta_k +v_k \alpha_k^\dagger,   \)
            Here, $u_k=\cosh\eta_k$ and $v_k=\sinh \eta_k$ with $\eta_k$  determined by $\tanh 2\eta_k=-\gamma_k/(1+K_A/J_A)$. Then the Hamiltonian is diagonal in terms of operator $\alpha_k$ and $\beta_k$ with the dispersion being
            \(  \hbar\omega_k=2SJ_A \sqrt{ E_g^2+ k^2  }, \;\;\;\text{with}\;\; E_g= \sqrt{(1+K_A/J_A)^2-1}.   \)
            For magnonic excitations with small $k$, the spectrum is given by
            \( \omega_k \approx Ak^2 +\Delta, \;\;\;\; \text{with}\;\; \Delta=\frac{2SJ_AE_g}{\hbar}, \;\; A= \frac{SJ_A}{\hbar E_g}.  \)
            
            The second-order effective interaction between the vortices can be obtained by employing a perturbative Schrieffer-Wolff transformation~\cite{bravyi2011schrieffer}: 
            \( \mathcal{V}=-\frac{i}{2\hbar} \lim_{\eta\rightarrow0^+}\int^\infty_0dt\, e^{-\eta t} [H^\prime(t), H^\prime], \)
where $H^\prime(t)$ is in the interaction picture. We  then have the following effective interaction between vortices:
\(  \mathcal{V}=-\frac{\lambda^2\mathcal{J}_A^2}{4\hbar} \left\{ \left[  \chi_\perp(-\omega_R) d_{1R}^\dagger+\chi_\perp(\omega_L)d_{1L} \right] (d_{2R}^\dagger +d_{2L}   )  +  \left[  \chi_\perp(-\omega_L) d_{2L}^\dagger+\chi(\omega_R)d_{2R} \right] (d_{1L}^\dagger +d_{1R}   )   \right\} +\text{h.c.},   \label{s12}  \)
where the transversal susceptibility $\chi_\perp(\omega)\equiv \chi_\perp(\omega, L)$ (we have dropped  $L$ for simplicity) is defined as
\( \al{ \chi_\perp(\omega) & =-i \int^\infty_0 dt\, e^{-i\omega t} [n^+(t, L), n^-(0,0)]  \\
                                  &= -\frac{i}{2S} \int^\infty_0 dt \, e^{-i\omega t} [a_L(t)-b_L^\dagger(t), a^\dagger_0 -b_0].   } \label{s13} \)
                                  Here $a_L$, $b_L$, $a_0$, $b_0$ are magnon operators on site $L$ and site 0 ($L$ is the length of the coupler). 
 We have used N{\'e}el  vector $\vb n\equiv (\vb S_A-\vb S_B)/2S$, and we have neglected the longitudinal susceptibility in the interaction \eqref{s12} since  it is suppressed by temperature~\cite{Daniel2013prx}. To evaluate Eq.~\eqref{s13}, we write the expression in terms of $\alpha_k, \beta_k$ in momentum space which eventually yields: 
\( \al{ \chi_\perp(\omega) & =-\frac{i}{2S} \int^\infty_0 dt\, e^{-i\omega t-\eta t} \int \frac{dk}{2\pi} e^{ikL} (u_k-v_k)^2  \left( e^{-i\omega_k t} -e^{i\omega_k t} \right) \\
    &= \frac{1}{2S}  \int \frac{dk}{2\pi} e^{ikL} (u_k-v_k)^2  \left( \frac{1}{\omega-\omega_k-i\eta} - \frac{1}{\omega+\omega_k-i\eta} \right),  } \label{s14}  \)
where 
\( (u_k-v_k)^2=\cosh 2\eta_k -\sinh 2\eta_k=  \frac{1-\tanh (2\eta_k)}{    \sqrt{ 1-\tanh^2 (2\eta_k)   }} = \frac{2SJ_A( 1+K_A/J_A+\gamma_k  )}{\hbar \omega_k}.  \)
We note that Eq.~\eqref{s14} can be converted into a complex integral. When $\omega>0$, the main contribution of the integral comes from the pole $k=ik_F$ with $k_F^2=(\Delta-\omega)/A$, which leads to: 
\( \chi_\perp(\omega>0) \approx - \frac{( u_{ik_F}-v_{ik_F}  )^2}{4SAk_F}  e^{-k_FL}= - \frac{\alpha_0}{Ak_F} e^{-k_FL}  .  \) 
Similarly, for $\omega<0$, the main contribution still comes from $k=ik_F$ with $k_F^2=(\Delta-|\omega|)/A$. Thus we have $\chi_\perp(\omega<0)=\chi_{\perp}(|\omega|>0)$. 
Here $\alpha_0$ is a dimensionless numeric factor:
\(  \alpha_0=  \frac{1}{4S} (u_{ik_F} -v_{ik_F}  )^2=\frac{1}{4S} \sqrt{ \frac{K_A/J_A +2+k_F^2/2}{ K_A/J_A-k_F^2/2    }  }  \approx \frac{1}{4S} \sqrt{1+ \frac{2J_A}{K_A}},   \)
which is typically of order one when we have a sizable $K_A$ (so we have a large magnon gap $\Delta>\omega_L, \omega_R$). Here we have assumed $k_F\ll 1$ which is also required if we need the vortex-vortex interaction to be sizable for $L\gg 1$ (recall that here  we have set the lattice spacing $a=1$ for simplicity).

When $\omega_R=\omega_L=\Omega_0$ in the absence of magnetic field, the effective interaction can be simplified to
\( \al{   \mathcal{V} &=- \frac{\lambda^2\mathcal{J}_A^2\chi_{\perp}(\Omega_0)}{4\hbar} \left[ (d_{1R}^\dagger  +d_{1L}  )(d_{2R}^\dagger +d_{2L}) +( d_{2L}^\dagger +d_{2R} )(d^\dagger_{1L} +d_{1R})   \right] +\text{h.c.}  \\
      &=   \frac{\mathcal{J}_A^2 \chi_\perp (\Omega_0) }{2\hbar} \left( -\mathcal{X}_1\mathcal{X}_2+\mathcal{Y}_1\mathcal{Y}_2  \right)  ,  }  \)
where we have rewritten the effective potential in terms of the collective coordinates of the vortices.  Here we have focused on the case of $p=+1$, $\mathcal{N}=+1$. For polarity $p=\pm 1$ and topological charge $\mathcal{N}=\pm 1$, one can easily verify that the effective interaction between vortices is given by Eq. (10) in the main text. 

\begin{figure}
\centering\includegraphics[width=0.6\textwidth,trim=0.6cm 0cm 0cm 0.1cm]{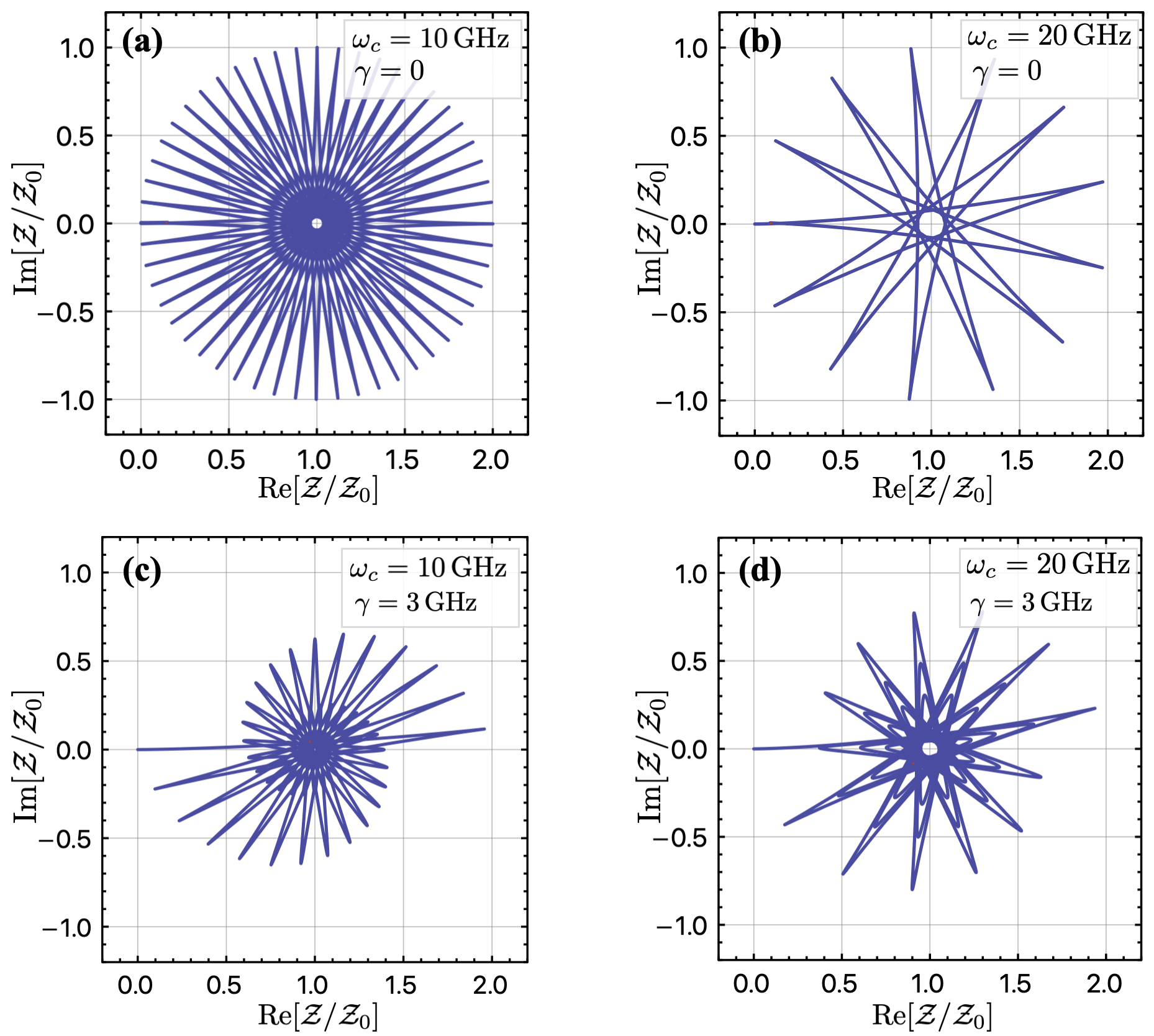}
\caption{Vortex trajectory. We plot the trajectories of the vortex with a current in the $y$ direction for (a) $\omega_c=10\, \text{GHz}$ and zero damping, (b) $\omega_c=20\, \text{GHz}$ and zero damping, (c) $\omega_c=10\, \text{GHz}$ and $\gamma=3\,\text{GHz}$, and (d) $\omega_c=20\, \text{GHz}$ and $\gamma=3\,\text{GHz}$.  }
\label{smf1}
\end{figure}

\subsection*{(ii) Trajectories of the vortex}
Here we discuss the trajectories of the vortex in the magnetic dot with the applied magnetic field and current. Following directly from the equation of motion for the vortex [Eq. (7) in the main text], we write down the following equation for the vortex trajectory: 
\( \partial_t^2\mathcal{Z}(t)-2\omega_ci \partial_t \mathcal{Z}(t)+2\gamma \partial_t \mathcal{Z}(t) +\Omega_0^2\mathcal{Z}(t) =\mathcal{Z}_0\Omega_0^2,  \)
where $\mathcal{Z}(t)=\mathcal{R}_x(t)+i\mathcal{R}_y(t)$ represents the vortex position and $\mathcal{Z}_0=2\omega_c(u_y-iu_x)/\Omega_0^2$ is the stationary position of the vortex with the applied magnetic field and in-plane current. Assuming that the vortex is static at the disk center at time $t=0$, the vortex trajectory is solve to be: 
\( \frac{\mathcal{Z}(t)}{\mathcal{Z}_0(t)}=1- \frac{\Omega_+ e^{-i\Omega_- t}  + \Omega_- e^{i\Omega_+ t}  }{\Omega_+ + \Omega_-} ,   \)
where $\Omega_{\pm}=\sqrt{\Omega_0^2 +(\omega_c+i\gamma)^2 } \pm (\omega_c+i\gamma) $. We show the trajectories of the vortex in Fig.~\ref{smf1} for different effective cyclotron frequencies (magnetic fields) and damping. Note that, in the absence of the damping $\gamma=0$, the vortex trajectory is reduced to $\mathcal{Z}(t)=\mathcal{Z}_0[1-(\omega_Re^{-i\omega_Lt} +\omega_Le^{i\omega_Rt} )/2\Omega]$ as we discussed in the main text.


\begin{thebibliography}{94}%
\makeatletter
\providecommand \@ifxundefined [1]{%
 \@ifx{#1\undefined}
}%
\providecommand \@ifnum [1]{%
 \ifnum #1\expandafter \@firstoftwo
 \else \expandafter \@secondoftwo
 \fi
}%
\providecommand \@ifx [1]{%
 \ifx #1\expandafter \@firstoftwo
 \else \expandafter \@secondoftwo
 \fi
}%
\providecommand \natexlab [1]{#1}%
\providecommand \enquote  [1]{``#1''}%
\providecommand \bibnamefont  [1]{#1}%
\providecommand \bibfnamefont [1]{#1}%
\providecommand \citenamefont [1]{#1}%
\providecommand \href@noop [0]{\@secondoftwo}%
\providecommand \href [0]{\begingroup \@sanitize@url \@href}%
\providecommand \@href[1]{\@@startlink{#1}\@@href}%
\providecommand \@@href[1]{\endgroup#1\@@endlink}%
\providecommand \@sanitize@url [0]{\catcode `\\12\catcode `\$12\catcode
  `\&12\catcode `\#12\catcode `\^12\catcode `\_12\catcode `\%12\relax}%
\providecommand \@@startlink[1]{}%
\providecommand \@@endlink[0]{}%
\providecommand \url  [0]{\begingroup\@sanitize@url \@url }%
\providecommand \@url [1]{\endgroup\@href {#1}{\urlprefix }}%
\providecommand \urlprefix  [0]{URL }%
\providecommand \Eprint [0]{\href }%
\providecommand \doibase [0]{https://doi.org/}%
\providecommand \selectlanguage [0]{\@gobble}%
\providecommand \bibinfo  [0]{\@secondoftwo}%
\providecommand \bibfield  [0]{\@secondoftwo}%
\providecommand \translation [1]{[#1]}%
\providecommand \BibitemOpen [0]{}%
\providecommand \bibitemStop [0]{}%
\providecommand \bibitemNoStop [0]{.\EOS\space}%
\providecommand \EOS [0]{\spacefactor3000\relax}%
\providecommand \BibitemShut  [1]{\csname bibitem#1\endcsname}%
\let\auto@bib@innerbib\@empty
\bibitem [{\citenamefont {Parkin}\ \emph {et~al.}(2008)\citenamefont {Parkin},
  \citenamefont {Hayashi},\ and\ \citenamefont {Thomas}}]{parkin2008magnetic}%
  \BibitemOpen
  \bibfield  {author} {\bibinfo {author} {\bibfnamefont {S.~S.}\ \bibnamefont
  {Parkin}}, \bibinfo {author} {\bibfnamefont {M.}~\bibnamefont {Hayashi}},\
  and\ \bibinfo {author} {\bibfnamefont {L.}~\bibnamefont {Thomas}},\
  }\bibfield  {title} {\bibinfo {title} {Magnetic domain-wall racetrack
  memory},\ }\href@noop {} {\bibfield  {journal} {\bibinfo  {journal}
  {Science}\ }\textbf {\bibinfo {volume} {320}},\ \bibinfo {pages} {190}
  (\bibinfo {year} {2008})}\BibitemShut {NoStop}%
\bibitem [{\citenamefont {Zang}\ \emph {et~al.}(2018)\citenamefont {Zang},
  \citenamefont {Cros},\ and\ \citenamefont {Hoffmann}}]{TopologyinMagnetism}%
  \BibitemOpen
  \bibinfo {editor} {\bibfnamefont {J.}~\bibnamefont {Zang}}, \bibinfo {editor}
  {\bibfnamefont {V.}~\bibnamefont {Cros}},\ and\ \bibinfo {editor}
  {\bibfnamefont {A.}~\bibnamefont {Hoffmann}},\ eds.,\ \href@noop {} {\emph
  {\bibinfo {title} {Topology in Magnetism}}}\ (\bibinfo  {publisher} {Springer
  International Publishing},\ \bibinfo {year} {2018})\BibitemShut {NoStop}%
\bibitem [{\citenamefont {Tserkovnyak}\ and\ \citenamefont
  {Xiao}(2018)}]{PhysRevLett.121.127701}%
  \BibitemOpen
  \bibfield  {author} {\bibinfo {author} {\bibfnamefont {Y.}~\bibnamefont
  {Tserkovnyak}}\ and\ \bibinfo {author} {\bibfnamefont {J.}~\bibnamefont
  {Xiao}},\ }\bibfield  {title} {\bibinfo {title} {Energy storage via
  topological spin textures},\ }\href@noop {} {\bibfield  {journal} {\bibinfo
  {journal} {Phys. Rev. Lett.}\ }\textbf {\bibinfo {volume} {121}},\ \bibinfo
  {pages} {127701} (\bibinfo {year} {2018})}\BibitemShut {NoStop}%
\bibitem [{\citenamefont {Liu}\ \emph {et~al.}(2018)\citenamefont {Liu},
  \citenamefont {Lake},\ and\ \citenamefont {Zang}}]{liu2018binding}%
  \BibitemOpen
  \bibfield  {author} {\bibinfo {author} {\bibfnamefont {Y.}~\bibnamefont
  {Liu}}, \bibinfo {author} {\bibfnamefont {R.~K.}\ \bibnamefont {Lake}},\ and\
  \bibinfo {author} {\bibfnamefont {J.}~\bibnamefont {Zang}},\ }\bibfield
  {title} {\bibinfo {title} {Binding a hopfion in a chiral magnet nanodisk},\
  }\href@noop {} {\bibfield  {journal} {\bibinfo  {journal} {Phys. Rev. B}\
  }\textbf {\bibinfo {volume} {98}},\ \bibinfo {pages} {174437} (\bibinfo
  {year} {2018})}\BibitemShut {NoStop}%
\bibitem [{\citenamefont {Tang}\ \emph {et~al.}(2021)\citenamefont {Tang},
  \citenamefont {Wu}, \citenamefont {Wang}, \citenamefont {Kong}, \citenamefont
  {Lv}, \citenamefont {Wei}, \citenamefont {Zang}, \citenamefont {Tian},\ and\
  \citenamefont {Du}}]{tang2021magnetic}%
  \BibitemOpen
  \bibfield  {author} {\bibinfo {author} {\bibfnamefont {J.}~\bibnamefont
  {Tang}}, \bibinfo {author} {\bibfnamefont {Y.}~\bibnamefont {Wu}}, \bibinfo
  {author} {\bibfnamefont {W.}~\bibnamefont {Wang}}, \bibinfo {author}
  {\bibfnamefont {L.}~\bibnamefont {Kong}}, \bibinfo {author} {\bibfnamefont
  {B.}~\bibnamefont {Lv}}, \bibinfo {author} {\bibfnamefont {W.}~\bibnamefont
  {Wei}}, \bibinfo {author} {\bibfnamefont {J.}~\bibnamefont {Zang}}, \bibinfo
  {author} {\bibfnamefont {M.}~\bibnamefont {Tian}},\ and\ \bibinfo {author}
  {\bibfnamefont {H.}~\bibnamefont {Du}},\ }\bibfield  {title} {\bibinfo
  {title} {Magnetic skyrmion bundles and their current-driven dynamics},\
  }\href@noop {} {\bibfield  {journal} {\bibinfo  {journal} {Nat.
  Nanotechnol.}\ }\textbf {\bibinfo {volume} {16}},\ \bibinfo {pages} {1086}
  (\bibinfo {year} {2021})}\BibitemShut {NoStop}%
\bibitem [{\citenamefont {Zarzuela}\ \emph {et~al.}(2020)\citenamefont
  {Zarzuela}, \citenamefont {Bharadwaj}, \citenamefont {Kim}, \citenamefont
  {Sinova},\ and\ \citenamefont {Everschor-Sitte}}]{zarzuela2020stability}%
  \BibitemOpen
  \bibfield  {author} {\bibinfo {author} {\bibfnamefont {R.}~\bibnamefont
  {Zarzuela}}, \bibinfo {author} {\bibfnamefont {V.~K.}\ \bibnamefont
  {Bharadwaj}}, \bibinfo {author} {\bibfnamefont {K.-W.}\ \bibnamefont {Kim}},
  \bibinfo {author} {\bibfnamefont {J.}~\bibnamefont {Sinova}},\ and\ \bibinfo
  {author} {\bibfnamefont {K.}~\bibnamefont {Everschor-Sitte}},\ }\bibfield
  {title} {\bibinfo {title} {Stability and dynamics of in-plane skyrmions in
  collinear ferromagnets},\ }\href@noop {} {\bibfield  {journal} {\bibinfo
  {journal} {Phys. Rev. B}\ }\textbf {\bibinfo {volume} {101}},\ \bibinfo
  {pages} {054405} (\bibinfo {year} {2020})}\BibitemShut {NoStop}%
\bibitem [{\citenamefont {Cheng}\ \emph {et~al.}(2019)\citenamefont {Cheng},
  \citenamefont {Li}, \citenamefont {Sapkota}, \citenamefont {Rai},
  \citenamefont {Pokhrel}, \citenamefont {Mewes}, \citenamefont {Mewes},
  \citenamefont {Xiao}, \citenamefont {De~Graef},\ and\ \citenamefont
  {Sokalski}}]{cheng2019magnetic}%
  \BibitemOpen
  \bibfield  {author} {\bibinfo {author} {\bibfnamefont {R.}~\bibnamefont
  {Cheng}}, \bibinfo {author} {\bibfnamefont {M.}~\bibnamefont {Li}}, \bibinfo
  {author} {\bibfnamefont {A.}~\bibnamefont {Sapkota}}, \bibinfo {author}
  {\bibfnamefont {A.}~\bibnamefont {Rai}}, \bibinfo {author} {\bibfnamefont
  {A.}~\bibnamefont {Pokhrel}}, \bibinfo {author} {\bibfnamefont
  {T.}~\bibnamefont {Mewes}}, \bibinfo {author} {\bibfnamefont
  {C.}~\bibnamefont {Mewes}}, \bibinfo {author} {\bibfnamefont
  {D.}~\bibnamefont {Xiao}}, \bibinfo {author} {\bibfnamefont {M.}~\bibnamefont
  {De~Graef}},\ and\ \bibinfo {author} {\bibfnamefont {V.}~\bibnamefont
  {Sokalski}},\ }\bibfield  {title} {\bibinfo {title} {Magnetic domain wall
  skyrmions},\ }\href@noop {} {\bibfield  {journal} {\bibinfo  {journal} {Phys.
  Rev. B}\ }\textbf {\bibinfo {volume} {99}},\ \bibinfo {pages} {184412}
  (\bibinfo {year} {2019})}\BibitemShut {NoStop}%
\bibitem [{\citenamefont {Psaroudaki}\ \emph {et~al.}(2023)\citenamefont
  {Psaroudaki}, \citenamefont {Peraticos},\ and\ \citenamefont
  {Panagopoulos}}]{psaroudaki2023skyrmion}%
  \BibitemOpen
  \bibfield  {author} {\bibinfo {author} {\bibfnamefont {C.}~\bibnamefont
  {Psaroudaki}}, \bibinfo {author} {\bibfnamefont {E.}~\bibnamefont
  {Peraticos}},\ and\ \bibinfo {author} {\bibfnamefont {C.}~\bibnamefont
  {Panagopoulos}},\ }\bibfield  {title} {\bibinfo {title} {Skyrmion qubits:
  Challenges for future quantum computing applications},\ }\href@noop {}
  {\bibfield  {journal} {\bibinfo  {journal} {Appl. Phys. Lett.}\ }\textbf
  {\bibinfo {volume} {123}} (\bibinfo {year} {2023})}\BibitemShut {NoStop}%
\bibitem [{\citenamefont {Psaroudaki}\ and\ \citenamefont
  {Panagopoulos}(2021)}]{christina_prl_2021}%
  \BibitemOpen
  \bibfield  {author} {\bibinfo {author} {\bibfnamefont {C.}~\bibnamefont
  {Psaroudaki}}\ and\ \bibinfo {author} {\bibfnamefont {C.}~\bibnamefont
  {Panagopoulos}},\ }\bibfield  {title} {\bibinfo {title} {Skyrmion qubits: A
  new class of quantum logic elements based on nanoscale magnetization},\
  }\href@noop {} {\bibfield  {journal} {\bibinfo  {journal} {Phys. Rev. Lett.}\
  }\textbf {\bibinfo {volume} {127}},\ \bibinfo {pages} {067201} (\bibinfo
  {year} {2021})}\BibitemShut {NoStop}%
\bibitem [{\citenamefont {Lan}\ \emph {et~al.}(2015)\citenamefont {Lan},
  \citenamefont {Yu}, \citenamefont {Wu}, \citenamefont {Xiao} \emph
  {et~al.}}]{lan2015spin}%
  \BibitemOpen
  \bibfield  {author} {\bibinfo {author} {\bibfnamefont {J.}~\bibnamefont
  {Lan}}, \bibinfo {author} {\bibfnamefont {W.}~\bibnamefont {Yu}}, \bibinfo
  {author} {\bibfnamefont {R.}~\bibnamefont {Wu}}, \bibinfo {author}
  {\bibfnamefont {J.}~\bibnamefont {Xiao}}, \emph {et~al.},\ }\bibfield
  {title} {\bibinfo {title} {Spin-wave diode},\ }\href@noop {} {\bibfield
  {journal} {\bibinfo  {journal} {Phys. Rev. X}\ }\textbf {\bibinfo {volume}
  {5}},\ \bibinfo {pages} {041049} (\bibinfo {year} {2015})}\BibitemShut
  {NoStop}%
\bibitem [{\citenamefont {Zou}\ \emph {et~al.}(2023{\natexlab{a}})\citenamefont
  {Zou}, \citenamefont {Bosco}, \citenamefont {Pal}, \citenamefont {Parkin},
  \citenamefont {Klinovaja},\ and\ \citenamefont {Loss}}]{Zou2023prr}%
  \BibitemOpen
  \bibfield  {author} {\bibinfo {author} {\bibfnamefont {J.}~\bibnamefont
  {Zou}}, \bibinfo {author} {\bibfnamefont {S.}~\bibnamefont {Bosco}}, \bibinfo
  {author} {\bibfnamefont {B.}~\bibnamefont {Pal}}, \bibinfo {author}
  {\bibfnamefont {S.~S.~P.}\ \bibnamefont {Parkin}}, \bibinfo {author}
  {\bibfnamefont {J.}~\bibnamefont {Klinovaja}},\ and\ \bibinfo {author}
  {\bibfnamefont {D.}~\bibnamefont {Loss}},\ }\bibfield  {title} {\bibinfo
  {title} {Quantum computing on magnetic racetracks with flying domain wall
  qubits},\ }\href@noop {} {\bibfield  {journal} {\bibinfo  {journal} {Phys.
  Rev. Res.}\ }\textbf {\bibinfo {volume} {5}},\ \bibinfo {pages} {033166}
  (\bibinfo {year} {2023}{\natexlab{a}})}\BibitemShut {NoStop}%
\bibitem [{\citenamefont {Jin}\ \emph {et~al.}(2023)\citenamefont {Jin},
  \citenamefont {Yao}, \citenamefont {Wang}, \citenamefont {Yuan},
  \citenamefont {Zeng}, \citenamefont {Wang}, \citenamefont {Cao},\ and\
  \citenamefont {Yan}}]{jin2023nonlinear}%
  \BibitemOpen
  \bibfield  {author} {\bibinfo {author} {\bibfnamefont {Z.}~\bibnamefont
  {Jin}}, \bibinfo {author} {\bibfnamefont {X.}~\bibnamefont {Yao}}, \bibinfo
  {author} {\bibfnamefont {Z.}~\bibnamefont {Wang}}, \bibinfo {author}
  {\bibfnamefont {H.}~\bibnamefont {Yuan}}, \bibinfo {author} {\bibfnamefont
  {Z.}~\bibnamefont {Zeng}}, \bibinfo {author} {\bibfnamefont {W.}~\bibnamefont
  {Wang}}, \bibinfo {author} {\bibfnamefont {Y.}~\bibnamefont {Cao}},\ and\
  \bibinfo {author} {\bibfnamefont {P.}~\bibnamefont {Yan}},\ }\bibfield
  {title} {\bibinfo {title} {Nonlinear topological magnon spin hall effect},\
  }\href@noop {} {\bibfield  {journal} {\bibinfo  {journal} {Phys. Rev. Lett.}\
  }\textbf {\bibinfo {volume} {131}},\ \bibinfo {pages} {166704} (\bibinfo
  {year} {2023})}\BibitemShut {NoStop}%
\bibitem [{\citenamefont {Xia}\ \emph {et~al.}(2023)\citenamefont {Xia},
  \citenamefont {Zhang}, \citenamefont {Liu}, \citenamefont {Zhou},\ and\
  \citenamefont {Ezawa}}]{xia2023universal}%
  \BibitemOpen
  \bibfield  {author} {\bibinfo {author} {\bibfnamefont {J.}~\bibnamefont
  {Xia}}, \bibinfo {author} {\bibfnamefont {X.}~\bibnamefont {Zhang}}, \bibinfo
  {author} {\bibfnamefont {X.}~\bibnamefont {Liu}}, \bibinfo {author}
  {\bibfnamefont {Y.}~\bibnamefont {Zhou}},\ and\ \bibinfo {author}
  {\bibfnamefont {M.}~\bibnamefont {Ezawa}},\ }\bibfield  {title} {\bibinfo
  {title} {Universal quantum computation based on nanoscale skyrmion helicity
  qubits in frustrated magnets},\ }\href@noop {} {\bibfield  {journal}
  {\bibinfo  {journal} {Phys. Rev. Lett.}\ }\textbf {\bibinfo {volume} {130}},\
  \bibinfo {pages} {106701} (\bibinfo {year} {2023})}\BibitemShut {NoStop}%
\bibitem [{\citenamefont {Daniels}\ \emph {et~al.}(2019)\citenamefont
  {Daniels}, \citenamefont {Yu}, \citenamefont {Cheng}, \citenamefont {Xiao},
  \citenamefont {Xiao} \emph {et~al.}}]{daniels2019topological}%
  \BibitemOpen
  \bibfield  {author} {\bibinfo {author} {\bibfnamefont {M.~W.}\ \bibnamefont
  {Daniels}}, \bibinfo {author} {\bibfnamefont {W.}~\bibnamefont {Yu}},
  \bibinfo {author} {\bibfnamefont {R.}~\bibnamefont {Cheng}}, \bibinfo
  {author} {\bibfnamefont {J.}~\bibnamefont {Xiao}}, \bibinfo {author}
  {\bibfnamefont {D.}~\bibnamefont {Xiao}}, \emph {et~al.},\ }\bibfield
  {title} {\bibinfo {title} {Topological spin hall effects and tunable skyrmion
  hall effects in uniaxial antiferromagnetic insulators},\ }\href@noop {}
  {\bibfield  {journal} {\bibinfo  {journal} {Phys. Rev. B}\ }\textbf {\bibinfo
  {volume} {99}},\ \bibinfo {pages} {224433} (\bibinfo {year}
  {2019})}\BibitemShut {NoStop}%
\bibitem [{\citenamefont {Takei}\ and\ \citenamefont
  {Mohseni}(2018)}]{PhysRevB.97.064401}%
  \BibitemOpen
  \bibfield  {author} {\bibinfo {author} {\bibfnamefont {S.}~\bibnamefont
  {Takei}}\ and\ \bibinfo {author} {\bibfnamefont {M.}~\bibnamefont
  {Mohseni}},\ }\bibfield  {title} {\bibinfo {title} {Quantum control of
  topological defects in magnetic systems},\ }\href@noop {} {\bibfield
  {journal} {\bibinfo  {journal} {Phys. Rev. B}\ }\textbf {\bibinfo {volume}
  {97}},\ \bibinfo {pages} {064401} (\bibinfo {year} {2018})}\BibitemShut
  {NoStop}%
\bibitem [{\citenamefont {Takei}\ \emph {et~al.}(2017)\citenamefont {Takei},
  \citenamefont {Tserkovnyak},\ and\ \citenamefont
  {Mohseni}}]{PhysRevB.95.144402}%
  \BibitemOpen
  \bibfield  {author} {\bibinfo {author} {\bibfnamefont {S.}~\bibnamefont
  {Takei}}, \bibinfo {author} {\bibfnamefont {Y.}~\bibnamefont {Tserkovnyak}},\
  and\ \bibinfo {author} {\bibfnamefont {M.}~\bibnamefont {Mohseni}},\
  }\bibfield  {title} {\bibinfo {title} {Spin superfluid josephson quantum
  devices},\ }\href@noop {} {\bibfield  {journal} {\bibinfo  {journal} {Phys.
  Rev. B}\ }\textbf {\bibinfo {volume} {95}},\ \bibinfo {pages} {144402}
  (\bibinfo {year} {2017})}\BibitemShut {NoStop}%
\bibitem [{\citenamefont {Wang}\ \emph {et~al.}(2018)\citenamefont {Wang},
  \citenamefont {Yuan},\ and\ \citenamefont {Wang}}]{wang2018theory}%
  \BibitemOpen
  \bibfield  {author} {\bibinfo {author} {\bibfnamefont {X.}~\bibnamefont
  {Wang}}, \bibinfo {author} {\bibfnamefont {H.}~\bibnamefont {Yuan}},\ and\
  \bibinfo {author} {\bibfnamefont {X.}~\bibnamefont {Wang}},\ }\bibfield
  {title} {\bibinfo {title} {A theory on skyrmion size},\ }\href@noop {}
  {\bibfield  {journal} {\bibinfo  {journal} {Communications Physics}\ }\textbf
  {\bibinfo {volume} {1}},\ \bibinfo {pages} {31} (\bibinfo {year}
  {2018})}\BibitemShut {NoStop}%
\bibitem [{\citenamefont {Zou}\ \emph {et~al.}(2020)\citenamefont {Zou},
  \citenamefont {Zhang},\ and\ \citenamefont {Tserkovnyak}}]{jiprl2020}%
  \BibitemOpen
  \bibfield  {author} {\bibinfo {author} {\bibfnamefont {J.}~\bibnamefont
  {Zou}}, \bibinfo {author} {\bibfnamefont {S.}~\bibnamefont {Zhang}},\ and\
  \bibinfo {author} {\bibfnamefont {Y.}~\bibnamefont {Tserkovnyak}},\
  }\bibfield  {title} {\bibinfo {title} {Topological transport of deconfined
  hedgehogs in magnets},\ }\href@noop {} {\bibfield  {journal} {\bibinfo
  {journal} {Phys. Rev. Lett.}\ }\textbf {\bibinfo {volume} {125}},\ \bibinfo
  {pages} {267201} (\bibinfo {year} {2020})}\BibitemShut {NoStop}%
\bibitem [{\citenamefont {Wang}\ \emph {et~al.}(2019)\citenamefont {Wang},
  \citenamefont {Qaiumzadeh},\ and\ \citenamefont
  {Brataas}}]{PhysRevLett.123.147203}%
  \BibitemOpen
  \bibfield  {author} {\bibinfo {author} {\bibfnamefont {X.~S.}\ \bibnamefont
  {Wang}}, \bibinfo {author} {\bibfnamefont {A.}~\bibnamefont {Qaiumzadeh}},\
  and\ \bibinfo {author} {\bibfnamefont {A.}~\bibnamefont {Brataas}},\
  }\bibfield  {title} {\bibinfo {title} {Current-driven dynamics of magnetic
  hopfions},\ }\href@noop {} {\bibfield  {journal} {\bibinfo  {journal} {Phys.
  Rev. Lett.}\ }\textbf {\bibinfo {volume} {123}},\ \bibinfo {pages} {147203}
  (\bibinfo {year} {2019})}\BibitemShut {NoStop}%
\bibitem [{\citenamefont {Liu}\ \emph {et~al.}(2020)\citenamefont {Liu},
  \citenamefont {Hou}, \citenamefont {Han},\ and\ \citenamefont
  {Zang}}]{Zangprl}%
  \BibitemOpen
  \bibfield  {author} {\bibinfo {author} {\bibfnamefont {Y.}~\bibnamefont
  {Liu}}, \bibinfo {author} {\bibfnamefont {W.}~\bibnamefont {Hou}}, \bibinfo
  {author} {\bibfnamefont {X.}~\bibnamefont {Han}},\ and\ \bibinfo {author}
  {\bibfnamefont {J.}~\bibnamefont {Zang}},\ }\bibfield  {title} {\bibinfo
  {title} {Three-dimensional dynamics of a magnetic hopfion driven by spin
  transfer torque},\ }\href@noop {} {\bibfield  {journal} {\bibinfo  {journal}
  {Phys. Rev. Lett.}\ }\textbf {\bibinfo {volume} {124}},\ \bibinfo {pages}
  {127204} (\bibinfo {year} {2020})}\BibitemShut {NoStop}%
\bibitem [{\citenamefont {Tserkovnyak}\ \emph {et~al.}(2020)\citenamefont
  {Tserkovnyak}, \citenamefont {Zou}, \citenamefont {Kim},\ and\ \citenamefont
  {Takei}}]{Yqwinding}%
  \BibitemOpen
  \bibfield  {author} {\bibinfo {author} {\bibfnamefont {Y.}~\bibnamefont
  {Tserkovnyak}}, \bibinfo {author} {\bibfnamefont {J.}~\bibnamefont {Zou}},
  \bibinfo {author} {\bibfnamefont {S.~K.}\ \bibnamefont {Kim}},\ and\ \bibinfo
  {author} {\bibfnamefont {S.}~\bibnamefont {Takei}},\ }\bibfield  {title}
  {\bibinfo {title} {Quantum hydrodynamics of spin winding},\ }\href@noop {}
  {\bibfield  {journal} {\bibinfo  {journal} {Phys. Rev. B}\ }\textbf {\bibinfo
  {volume} {102}},\ \bibinfo {pages} {224433} (\bibinfo {year}
  {2020})}\BibitemShut {NoStop}%
\bibitem [{\citenamefont {Zou}\ \emph {et~al.}(2019)\citenamefont {Zou},
  \citenamefont {Kim},\ and\ \citenamefont {Tserkovnyak}}]{jivortex}%
  \BibitemOpen
  \bibfield  {author} {\bibinfo {author} {\bibfnamefont {J.}~\bibnamefont
  {Zou}}, \bibinfo {author} {\bibfnamefont {S.~K.}\ \bibnamefont {Kim}},\ and\
  \bibinfo {author} {\bibfnamefont {Y.}~\bibnamefont {Tserkovnyak}},\
  }\bibfield  {title} {\bibinfo {title} {Topological transport of vorticity in
  heisenberg magnets},\ }\href@noop {} {\bibfield  {journal} {\bibinfo
  {journal} {Phys. Rev. B}\ }\textbf {\bibinfo {volume} {99}},\ \bibinfo
  {pages} {180402} (\bibinfo {year} {2019})}\BibitemShut {NoStop}%
\bibitem [{\citenamefont {Zarzuela}\ and\ \citenamefont
  {Sinova}(2023)}]{zarzuela2023spin}%
  \BibitemOpen
  \bibfield  {author} {\bibinfo {author} {\bibfnamefont {R.}~\bibnamefont
  {Zarzuela}}\ and\ \bibinfo {author} {\bibfnamefont {J.}~\bibnamefont
  {Sinova}},\ }\bibfield  {title} {\bibinfo {title} {Spin-transfer and
  topological hall effects in itinerant frustrated magnets},\ }\href@noop {}
  {\bibfield  {journal} {\bibinfo  {journal} {Phys. Rev. B}\ }\textbf {\bibinfo
  {volume} {108}},\ \bibinfo {pages} {134402} (\bibinfo {year}
  {2023})}\BibitemShut {NoStop}%
\bibitem [{\citenamefont {Yamane}\ \emph {et~al.}(2019)\citenamefont {Yamane},
  \citenamefont {Gomonay},\ and\ \citenamefont {Sinova}}]{yamane2019dynamics}%
  \BibitemOpen
  \bibfield  {author} {\bibinfo {author} {\bibfnamefont {Y.}~\bibnamefont
  {Yamane}}, \bibinfo {author} {\bibfnamefont {O.}~\bibnamefont {Gomonay}},\
  and\ \bibinfo {author} {\bibfnamefont {J.}~\bibnamefont {Sinova}},\
  }\bibfield  {title} {\bibinfo {title} {Dynamics of noncollinear
  antiferromagnetic textures driven by spin current injection},\ }\href@noop {}
  {\bibfield  {journal} {\bibinfo  {journal} {Phys. Rev. B}\ }\textbf {\bibinfo
  {volume} {100}},\ \bibinfo {pages} {054415} (\bibinfo {year}
  {2019})}\BibitemShut {NoStop}%
\bibitem [{\citenamefont {Dao}\ \emph {et~al.}(2023)\citenamefont {Dao},
  \citenamefont {Zou}, \citenamefont {Kleinherbers},\ and\ \citenamefont
  {Tserkovnyak}}]{dao2023topological}%
  \BibitemOpen
  \bibfield  {author} {\bibinfo {author} {\bibfnamefont {C.}~\bibnamefont
  {Dao}}, \bibinfo {author} {\bibfnamefont {J.}~\bibnamefont {Zou}}, \bibinfo
  {author} {\bibfnamefont {E.}~\bibnamefont {Kleinherbers}},\ and\ \bibinfo
  {author} {\bibfnamefont {Y.}~\bibnamefont {Tserkovnyak}},\ }\bibfield
  {title} {\bibinfo {title} {Topological transport of vorticity on curved
  magnetic membranes},\ }\href@noop {} {\bibfield  {journal} {\bibinfo
  {journal} {arXiv:2311.00323}\ } (\bibinfo {year} {2023})}\BibitemShut
  {NoStop}%
\bibitem [{\citenamefont {Shinjo}\ \emph {et~al.}(2000)\citenamefont {Shinjo},
  \citenamefont {Okuno}, \citenamefont {Hassdorf}, \citenamefont {Shigeto},\
  and\ \citenamefont {Ono}}]{shinjo2000magnetic}%
  \BibitemOpen
  \bibfield  {author} {\bibinfo {author} {\bibfnamefont {T.}~\bibnamefont
  {Shinjo}}, \bibinfo {author} {\bibfnamefont {T.}~\bibnamefont {Okuno}},
  \bibinfo {author} {\bibfnamefont {R.}~\bibnamefont {Hassdorf}}, \bibinfo
  {author} {\bibfnamefont {K.}~\bibnamefont {Shigeto}},\ and\ \bibinfo {author}
  {\bibfnamefont {T.}~\bibnamefont {Ono}},\ }\bibfield  {title} {\bibinfo
  {title} {Magnetic vortex core observation in circular dots of permalloy},\
  }\href@noop {} {\bibfield  {journal} {\bibinfo  {journal} {Science}\ }\textbf
  {\bibinfo {volume} {289}},\ \bibinfo {pages} {930} (\bibinfo {year}
  {2000})}\BibitemShut {NoStop}%
\bibitem [{\citenamefont {Wachowiak}\ \emph {et~al.}(2002)\citenamefont
  {Wachowiak}, \citenamefont {Wiebe}, \citenamefont {Bode}, \citenamefont
  {Pietzsch}, \citenamefont {Morgenstern},\ and\ \citenamefont
  {Wiesendanger}}]{wachowiak2002direct}%
  \BibitemOpen
  \bibfield  {author} {\bibinfo {author} {\bibfnamefont {A.}~\bibnamefont
  {Wachowiak}}, \bibinfo {author} {\bibfnamefont {J.}~\bibnamefont {Wiebe}},
  \bibinfo {author} {\bibfnamefont {M.}~\bibnamefont {Bode}}, \bibinfo {author}
  {\bibfnamefont {O.}~\bibnamefont {Pietzsch}}, \bibinfo {author}
  {\bibfnamefont {M.}~\bibnamefont {Morgenstern}},\ and\ \bibinfo {author}
  {\bibfnamefont {R.}~\bibnamefont {Wiesendanger}},\ }\bibfield  {title}
  {\bibinfo {title} {Direct observation of internal spin structure of magnetic
  vortex cores},\ }\href@noop {} {\bibfield  {journal} {\bibinfo  {journal}
  {Science}\ }\textbf {\bibinfo {volume} {298}},\ \bibinfo {pages} {577}
  (\bibinfo {year} {2002})}\BibitemShut {NoStop}%
\bibitem [{\citenamefont {Dussaux}\ \emph {et~al.}(2010)\citenamefont
  {Dussaux}, \citenamefont {Georges}, \citenamefont {Grollier}, \citenamefont
  {Cros}, \citenamefont {Khvalkovskiy}, \citenamefont {Fukushima},
  \citenamefont {Konoto}, \citenamefont {Kubota}, \citenamefont {Yakushiji},
  \citenamefont {Yuasa} \emph {et~al.}}]{dussaux2010large}%
  \BibitemOpen
  \bibfield  {author} {\bibinfo {author} {\bibfnamefont {A.}~\bibnamefont
  {Dussaux}}, \bibinfo {author} {\bibfnamefont {B.}~\bibnamefont {Georges}},
  \bibinfo {author} {\bibfnamefont {J.}~\bibnamefont {Grollier}}, \bibinfo
  {author} {\bibfnamefont {V.}~\bibnamefont {Cros}}, \bibinfo {author}
  {\bibfnamefont {A.}~\bibnamefont {Khvalkovskiy}}, \bibinfo {author}
  {\bibfnamefont {A.}~\bibnamefont {Fukushima}}, \bibinfo {author}
  {\bibfnamefont {M.}~\bibnamefont {Konoto}}, \bibinfo {author} {\bibfnamefont
  {H.}~\bibnamefont {Kubota}}, \bibinfo {author} {\bibfnamefont
  {K.}~\bibnamefont {Yakushiji}}, \bibinfo {author} {\bibfnamefont
  {S.}~\bibnamefont {Yuasa}}, \emph {et~al.},\ }\bibfield  {title} {\bibinfo
  {title} {Large microwave generation from current-driven magnetic vortex
  oscillators in magnetic tunnel junctions},\ }\href@noop {} {\bibfield
  {journal} {\bibinfo  {journal} {Nat. Commun.}\ }\textbf {\bibinfo {volume}
  {1}},\ \bibinfo {pages} {1} (\bibinfo {year} {2010})}\BibitemShut {NoStop}%
\bibitem [{\citenamefont {Wintz}\ \emph {et~al.}(2016)\citenamefont {Wintz},
  \citenamefont {Tiberkevich}, \citenamefont {Weigand}, \citenamefont {Raabe},
  \citenamefont {Lindner}, \citenamefont {Erbe}, \citenamefont {Slavin},\ and\
  \citenamefont {Fassbender}}]{wintz2016magnetic}%
  \BibitemOpen
  \bibfield  {author} {\bibinfo {author} {\bibfnamefont {S.}~\bibnamefont
  {Wintz}}, \bibinfo {author} {\bibfnamefont {V.}~\bibnamefont {Tiberkevich}},
  \bibinfo {author} {\bibfnamefont {M.}~\bibnamefont {Weigand}}, \bibinfo
  {author} {\bibfnamefont {J.}~\bibnamefont {Raabe}}, \bibinfo {author}
  {\bibfnamefont {J.}~\bibnamefont {Lindner}}, \bibinfo {author} {\bibfnamefont
  {A.}~\bibnamefont {Erbe}}, \bibinfo {author} {\bibfnamefont {A.}~\bibnamefont
  {Slavin}},\ and\ \bibinfo {author} {\bibfnamefont {J.}~\bibnamefont
  {Fassbender}},\ }\bibfield  {title} {\bibinfo {title} {Magnetic vortex cores
  as tunable spin-wave emitters},\ }\href@noop {} {\bibfield  {journal}
  {\bibinfo  {journal} {Nat. Nanotechnol.}\ }\textbf {\bibinfo {volume} {11}},\
  \bibinfo {pages} {948} (\bibinfo {year} {2016})}\BibitemShut {NoStop}%
\bibitem [{\citenamefont {Choi}\ \emph {et~al.}(2007)\citenamefont {Choi},
  \citenamefont {Lee}, \citenamefont {Guslienko},\ and\ \citenamefont
  {Kim}}]{choi2007}%
  \BibitemOpen
  \bibfield  {author} {\bibinfo {author} {\bibfnamefont {S.}~\bibnamefont
  {Choi}}, \bibinfo {author} {\bibfnamefont {K.-S.}\ \bibnamefont {Lee}},
  \bibinfo {author} {\bibfnamefont {K.~Y.}\ \bibnamefont {Guslienko}},\ and\
  \bibinfo {author} {\bibfnamefont {S.-K.}\ \bibnamefont {Kim}},\ }\bibfield
  {title} {\bibinfo {title} {Strong radiation of spin waves by core reversal of
  a magnetic vortex and their wave behaviors in magnetic nanowire waveguides},\
  }\href@noop {} {\bibfield  {journal} {\bibinfo  {journal} {Phys. Rev. Lett.}\
  }\textbf {\bibinfo {volume} {98}},\ \bibinfo {pages} {087205} (\bibinfo
  {year} {2007})}\BibitemShut {NoStop}%
\bibitem [{\citenamefont {Yamada}\ \emph {et~al.}(2007)\citenamefont {Yamada},
  \citenamefont {Kasai}, \citenamefont {Nakatani}, \citenamefont {Kobayashi},
  \citenamefont {Kohno}, \citenamefont {Thiaville},\ and\ \citenamefont
  {Ono}}]{yamada2007electrical}%
  \BibitemOpen
  \bibfield  {author} {\bibinfo {author} {\bibfnamefont {K.}~\bibnamefont
  {Yamada}}, \bibinfo {author} {\bibfnamefont {S.}~\bibnamefont {Kasai}},
  \bibinfo {author} {\bibfnamefont {Y.}~\bibnamefont {Nakatani}}, \bibinfo
  {author} {\bibfnamefont {K.}~\bibnamefont {Kobayashi}}, \bibinfo {author}
  {\bibfnamefont {H.}~\bibnamefont {Kohno}}, \bibinfo {author} {\bibfnamefont
  {A.}~\bibnamefont {Thiaville}},\ and\ \bibinfo {author} {\bibfnamefont
  {T.}~\bibnamefont {Ono}},\ }\bibfield  {title} {\bibinfo {title} {Electrical
  switching of the vortex core in a magnetic disk},\ }\href@noop {} {\bibfield
  {journal} {\bibinfo  {journal} {Nat. Mater.}\ }\textbf {\bibinfo {volume}
  {6}},\ \bibinfo {pages} {270} (\bibinfo {year} {2007})}\BibitemShut {NoStop}%
\bibitem [{\citenamefont {Guslienko}\ \emph {et~al.}(2008)\citenamefont
  {Guslienko}, \citenamefont {Lee},\ and\ \citenamefont
  {Kim}}]{PhysRevLett.100.027203}%
  \BibitemOpen
  \bibfield  {author} {\bibinfo {author} {\bibfnamefont {K.~Y.}\ \bibnamefont
  {Guslienko}}, \bibinfo {author} {\bibfnamefont {K.-S.}\ \bibnamefont {Lee}},\
  and\ \bibinfo {author} {\bibfnamefont {S.-K.}\ \bibnamefont {Kim}},\
  }\bibfield  {title} {\bibinfo {title} {Dynamic origin of vortex core
  switching in soft magnetic nanodots},\ }\href@noop {} {\bibfield  {journal}
  {\bibinfo  {journal} {Phys. Rev. Lett.}\ }\textbf {\bibinfo {volume} {100}},\
  \bibinfo {pages} {027203} (\bibinfo {year} {2008})}\BibitemShut {NoStop}%
\bibitem [{\citenamefont {Pigeau}\ \emph {et~al.}(2010)\citenamefont {Pigeau},
  \citenamefont {De~Loubens}, \citenamefont {Klein}, \citenamefont {Riegler},
  \citenamefont {Lochner}, \citenamefont {Schmidt}, \citenamefont {Molenkamp},
  \citenamefont {Tiberkevich},\ and\ \citenamefont
  {Slavin}}]{pigeau2010frequency}%
  \BibitemOpen
  \bibfield  {author} {\bibinfo {author} {\bibfnamefont {B.}~\bibnamefont
  {Pigeau}}, \bibinfo {author} {\bibfnamefont {G.}~\bibnamefont {De~Loubens}},
  \bibinfo {author} {\bibfnamefont {O.}~\bibnamefont {Klein}}, \bibinfo
  {author} {\bibfnamefont {A.}~\bibnamefont {Riegler}}, \bibinfo {author}
  {\bibfnamefont {F.}~\bibnamefont {Lochner}}, \bibinfo {author} {\bibfnamefont
  {G.}~\bibnamefont {Schmidt}}, \bibinfo {author} {\bibfnamefont
  {L.}~\bibnamefont {Molenkamp}}, \bibinfo {author} {\bibfnamefont
  {V.}~\bibnamefont {Tiberkevich}},\ and\ \bibinfo {author} {\bibfnamefont
  {A.}~\bibnamefont {Slavin}},\ }\bibfield  {title} {\bibinfo {title} {A
  frequency-controlled magnetic vortex memory},\ }\href@noop {} {\bibfield
  {journal} {\bibinfo  {journal} {Appl. Phys. Lett.}\ }\textbf {\bibinfo
  {volume} {96}} (\bibinfo {year} {2010})}\BibitemShut {NoStop}%
\bibitem [{\citenamefont {Bohlens}\ \emph {et~al.}(2008)\citenamefont
  {Bohlens}, \citenamefont {Kr{\"u}ger}, \citenamefont {Drews}, \citenamefont
  {Bolte}, \citenamefont {Meier},\ and\ \citenamefont
  {Pfannkuche}}]{bohlens2008current}%
  \BibitemOpen
  \bibfield  {author} {\bibinfo {author} {\bibfnamefont {S.}~\bibnamefont
  {Bohlens}}, \bibinfo {author} {\bibfnamefont {B.}~\bibnamefont {Kr{\"u}ger}},
  \bibinfo {author} {\bibfnamefont {A.}~\bibnamefont {Drews}}, \bibinfo
  {author} {\bibfnamefont {M.}~\bibnamefont {Bolte}}, \bibinfo {author}
  {\bibfnamefont {G.}~\bibnamefont {Meier}},\ and\ \bibinfo {author}
  {\bibfnamefont {D.}~\bibnamefont {Pfannkuche}},\ }\bibfield  {title}
  {\bibinfo {title} {Current controlled random-access memory based on magnetic
  vortex handedness},\ }\href@noop {} {\bibfield  {journal} {\bibinfo
  {journal} {Appl. Phys. Lett.}\ }\textbf {\bibinfo {volume} {93}} (\bibinfo
  {year} {2008})}\BibitemShut {NoStop}%
\bibitem [{\citenamefont {Hrkac}\ \emph {et~al.}(2015)\citenamefont {Hrkac},
  \citenamefont {Keatley}, \citenamefont {Bryan},\ and\ \citenamefont
  {Butler}}]{hrkac2015magnetic}%
  \BibitemOpen
  \bibfield  {author} {\bibinfo {author} {\bibfnamefont {G.}~\bibnamefont
  {Hrkac}}, \bibinfo {author} {\bibfnamefont {P.~S.}\ \bibnamefont {Keatley}},
  \bibinfo {author} {\bibfnamefont {M.~T.}\ \bibnamefont {Bryan}},\ and\
  \bibinfo {author} {\bibfnamefont {K.}~\bibnamefont {Butler}},\ }\bibfield
  {title} {\bibinfo {title} {Magnetic vortex oscillators},\ }\href@noop {}
  {\bibfield  {journal} {\bibinfo  {journal} {J. Phys. D: Appl. Phys.}\
  }\textbf {\bibinfo {volume} {48}},\ \bibinfo {pages} {453001} (\bibinfo
  {year} {2015})}\BibitemShut {NoStop}%
\bibitem [{\citenamefont {Jung}\ \emph {et~al.}(2012)\citenamefont {Jung},
  \citenamefont {Choi}, \citenamefont {Lee}, \citenamefont {Han}, \citenamefont
  {Yu}, \citenamefont {Im}, \citenamefont {Fischer},\ and\ \citenamefont
  {Kim}}]{jung2012logic}%
  \BibitemOpen
  \bibfield  {author} {\bibinfo {author} {\bibfnamefont {H.}~\bibnamefont
  {Jung}}, \bibinfo {author} {\bibfnamefont {Y.-S.}\ \bibnamefont {Choi}},
  \bibinfo {author} {\bibfnamefont {K.-S.}\ \bibnamefont {Lee}}, \bibinfo
  {author} {\bibfnamefont {D.-S.}\ \bibnamefont {Han}}, \bibinfo {author}
  {\bibfnamefont {Y.-S.}\ \bibnamefont {Yu}}, \bibinfo {author} {\bibfnamefont
  {M.-Y.}\ \bibnamefont {Im}}, \bibinfo {author} {\bibfnamefont
  {P.}~\bibnamefont {Fischer}},\ and\ \bibinfo {author} {\bibfnamefont {S.-K.}\
  \bibnamefont {Kim}},\ }\bibfield  {title} {\bibinfo {title} {Logic operations
  based on magnetic-vortex-state networks},\ }\href@noop {} {\bibfield
  {journal} {\bibinfo  {journal} {Acs Nano}\ }\textbf {\bibinfo {volume} {6}},\
  \bibinfo {pages} {3712} (\bibinfo {year} {2012})}\BibitemShut {NoStop}%
\bibitem [{\citenamefont {Shreya}\ \emph {et~al.}(2022)\citenamefont {Shreya},
  \citenamefont {Zamani}, \citenamefont {Rezaeiyan}, \citenamefont {Ghanatian},
  \citenamefont {B{\"o}hnert}, \citenamefont {Jenkins}, \citenamefont
  {Ferreira}, \citenamefont {Farkhani},\ and\ \citenamefont
  {Moradi}}]{shreya2022memory}%
  \BibitemOpen
  \bibfield  {author} {\bibinfo {author} {\bibfnamefont {S.}~\bibnamefont
  {Shreya}}, \bibinfo {author} {\bibfnamefont {M.}~\bibnamefont {Zamani}},
  \bibinfo {author} {\bibfnamefont {Y.}~\bibnamefont {Rezaeiyan}}, \bibinfo
  {author} {\bibfnamefont {H.}~\bibnamefont {Ghanatian}}, \bibinfo {author}
  {\bibfnamefont {T.}~\bibnamefont {B{\"o}hnert}}, \bibinfo {author}
  {\bibfnamefont {A.~S.}\ \bibnamefont {Jenkins}}, \bibinfo {author}
  {\bibfnamefont {R.}~\bibnamefont {Ferreira}}, \bibinfo {author}
  {\bibfnamefont {H.}~\bibnamefont {Farkhani}},\ and\ \bibinfo {author}
  {\bibfnamefont {F.}~\bibnamefont {Moradi}},\ }\bibfield  {title} {\bibinfo
  {title} {Memory and communication-in-logic using vortex and precessional
  oscillations in a magnetic tunnel junction},\ }\href@noop {} {\bibfield
  {journal} {\bibinfo  {journal} {IEEE Magn. Lett.}\ }\textbf {\bibinfo
  {volume} {13}},\ \bibinfo {pages} {1} (\bibinfo {year} {2022})}\BibitemShut
  {NoStop}%
\bibitem [{\citenamefont {Omari}\ and\ \citenamefont
  {Hayward}(2014)}]{PhysRevApplied.2.044001}%
  \BibitemOpen
  \bibfield  {author} {\bibinfo {author} {\bibfnamefont {K.~A.}\ \bibnamefont
  {Omari}}\ and\ \bibinfo {author} {\bibfnamefont {T.~J.}\ \bibnamefont
  {Hayward}},\ }\bibfield  {title} {\bibinfo {title} {Chirality-based vortex
  domain-wall logic gates},\ }\href@noop {} {\bibfield  {journal} {\bibinfo
  {journal} {Phys. Rev. Appl.}\ }\textbf {\bibinfo {volume} {2}},\ \bibinfo
  {pages} {044001} (\bibinfo {year} {2014})}\BibitemShut {NoStop}%
\bibitem [{\citenamefont {Torrejon}\ \emph {et~al.}(2017)\citenamefont
  {Torrejon}, \citenamefont {Riou}, \citenamefont {Araujo}, \citenamefont
  {Tsunegi}, \citenamefont {Khalsa}, \citenamefont {Querlioz}, \citenamefont
  {Bortolotti}, \citenamefont {Cros}, \citenamefont {Yakushiji}, \citenamefont
  {Fukushima} \emph {et~al.}}]{torrejon2017neuromorphic}%
  \BibitemOpen
  \bibfield  {author} {\bibinfo {author} {\bibfnamefont {J.}~\bibnamefont
  {Torrejon}}, \bibinfo {author} {\bibfnamefont {M.}~\bibnamefont {Riou}},
  \bibinfo {author} {\bibfnamefont {F.~A.}\ \bibnamefont {Araujo}}, \bibinfo
  {author} {\bibfnamefont {S.}~\bibnamefont {Tsunegi}}, \bibinfo {author}
  {\bibfnamefont {G.}~\bibnamefont {Khalsa}}, \bibinfo {author} {\bibfnamefont
  {D.}~\bibnamefont {Querlioz}}, \bibinfo {author} {\bibfnamefont
  {P.}~\bibnamefont {Bortolotti}}, \bibinfo {author} {\bibfnamefont
  {V.}~\bibnamefont {Cros}}, \bibinfo {author} {\bibfnamefont {K.}~\bibnamefont
  {Yakushiji}}, \bibinfo {author} {\bibfnamefont {A.}~\bibnamefont
  {Fukushima}}, \emph {et~al.},\ }\bibfield  {title} {\bibinfo {title}
  {Neuromorphic computing with nanoscale spintronic oscillators},\ }\href@noop
  {} {\bibfield  {journal} {\bibinfo  {journal} {Nature}\ }\textbf {\bibinfo
  {volume} {547}},\ \bibinfo {pages} {428} (\bibinfo {year}
  {2017})}\BibitemShut {NoStop}%
\bibitem [{\citenamefont {Yun}\ \emph {et~al.}(2022)\citenamefont {Yun},
  \citenamefont {Wu}, \citenamefont {Liang}, \citenamefont {Yang},
  \citenamefont {Du}, \citenamefont {Liu}, \citenamefont {Han}, \citenamefont
  {Hou}, \citenamefont {Yang},\ and\ \citenamefont {Luo}}]{yun2022magnetic}%
  \BibitemOpen
  \bibfield  {author} {\bibinfo {author} {\bibfnamefont {C.}~\bibnamefont
  {Yun}}, \bibinfo {author} {\bibfnamefont {Y.}~\bibnamefont {Wu}}, \bibinfo
  {author} {\bibfnamefont {Z.}~\bibnamefont {Liang}}, \bibinfo {author}
  {\bibfnamefont {W.}~\bibnamefont {Yang}}, \bibinfo {author} {\bibfnamefont
  {H.}~\bibnamefont {Du}}, \bibinfo {author} {\bibfnamefont {S.}~\bibnamefont
  {Liu}}, \bibinfo {author} {\bibfnamefont {J.}~\bibnamefont {Han}}, \bibinfo
  {author} {\bibfnamefont {Y.}~\bibnamefont {Hou}}, \bibinfo {author}
  {\bibfnamefont {J.}~\bibnamefont {Yang}},\ and\ \bibinfo {author}
  {\bibfnamefont {Z.}~\bibnamefont {Luo}},\ }\bibfield  {title} {\bibinfo
  {title} {Magnetic anisotropy-controlled vortex nano-oscillator for
  neuromorphic computing},\ }\href@noop {} {\bibfield  {journal} {\bibinfo
  {journal} {Front. Phys.}\ }\textbf {\bibinfo {volume} {10}},\ \bibinfo
  {pages} {1019881} (\bibinfo {year} {2022})}\BibitemShut {NoStop}%
\bibitem [{\citenamefont {Shreya}\ \emph {et~al.}(2023)\citenamefont {Shreya},
  \citenamefont {Jenkins}, \citenamefont {Rezaeiyan}, \citenamefont {Li},
  \citenamefont {B{\"o}hnert}, \citenamefont {Benetti}, \citenamefont
  {Ferreira}, \citenamefont {Moradi},\ and\ \citenamefont
  {Farkhani}}]{shreya2023granular}%
  \BibitemOpen
  \bibfield  {author} {\bibinfo {author} {\bibfnamefont {S.}~\bibnamefont
  {Shreya}}, \bibinfo {author} {\bibfnamefont {A.}~\bibnamefont {Jenkins}},
  \bibinfo {author} {\bibfnamefont {Y.}~\bibnamefont {Rezaeiyan}}, \bibinfo
  {author} {\bibfnamefont {R.}~\bibnamefont {Li}}, \bibinfo {author}
  {\bibfnamefont {T.}~\bibnamefont {B{\"o}hnert}}, \bibinfo {author}
  {\bibfnamefont {L.}~\bibnamefont {Benetti}}, \bibinfo {author} {\bibfnamefont
  {R.}~\bibnamefont {Ferreira}}, \bibinfo {author} {\bibfnamefont
  {F.}~\bibnamefont {Moradi}},\ and\ \bibinfo {author} {\bibfnamefont
  {H.}~\bibnamefont {Farkhani}},\ }\bibfield  {title} {\bibinfo {title}
  {Granular vortex spin-torque nano oscillator for reservoir computing},\
  }\href@noop {} {\bibfield  {journal} {\bibinfo  {journal} {Sci. Rep.}\
  }\textbf {\bibinfo {volume} {13}},\ \bibinfo {pages} {16722} (\bibinfo {year}
  {2023})}\BibitemShut {NoStop}%
\bibitem [{\citenamefont {Van~Waeyenberge}\ \emph {et~al.}(2006)\citenamefont
  {Van~Waeyenberge}, \citenamefont {Puzic}, \citenamefont {Stoll},
  \citenamefont {Chou}, \citenamefont {Tyliszczak}, \citenamefont {Hertel},
  \citenamefont {F{\"a}hnle}, \citenamefont {Br{\"u}ckl}, \citenamefont {Rott},
  \citenamefont {Reiss} \emph {et~al.}}]{van2006magnetic}%
  \BibitemOpen
  \bibfield  {author} {\bibinfo {author} {\bibfnamefont {B.}~\bibnamefont
  {Van~Waeyenberge}}, \bibinfo {author} {\bibfnamefont {A.}~\bibnamefont
  {Puzic}}, \bibinfo {author} {\bibfnamefont {H.}~\bibnamefont {Stoll}},
  \bibinfo {author} {\bibfnamefont {K.}~\bibnamefont {Chou}}, \bibinfo {author}
  {\bibfnamefont {T.}~\bibnamefont {Tyliszczak}}, \bibinfo {author}
  {\bibfnamefont {R.}~\bibnamefont {Hertel}}, \bibinfo {author} {\bibfnamefont
  {M.}~\bibnamefont {F{\"a}hnle}}, \bibinfo {author} {\bibfnamefont
  {H.}~\bibnamefont {Br{\"u}ckl}}, \bibinfo {author} {\bibfnamefont
  {K.}~\bibnamefont {Rott}}, \bibinfo {author} {\bibfnamefont {G.}~\bibnamefont
  {Reiss}}, \emph {et~al.},\ }\bibfield  {title} {\bibinfo {title} {Magnetic
  vortex core reversal by excitation with short bursts of an alternating
  field},\ }\href@noop {} {\bibfield  {journal} {\bibinfo  {journal} {Nature}\
  }\textbf {\bibinfo {volume} {444}},\ \bibinfo {pages} {461} (\bibinfo {year}
  {2006})}\BibitemShut {NoStop}%
\bibitem [{\citenamefont {Curcic}\ \emph {et~al.}(2008)\citenamefont {Curcic},
  \citenamefont {Van~Waeyenberge}, \citenamefont {Vansteenkiste}, \citenamefont
  {Weigand}, \citenamefont {Sackmann}, \citenamefont {Stoll}, \citenamefont
  {F\"ahnle}, \citenamefont {Tyliszczak}, \citenamefont {Woltersdorf},
  \citenamefont {Back},\ and\ \citenamefont
  {Sch\"utz}}]{PhysRevLett.101.197204}%
  \BibitemOpen
  \bibfield  {author} {\bibinfo {author} {\bibfnamefont {M.}~\bibnamefont
  {Curcic}}, \bibinfo {author} {\bibfnamefont {B.}~\bibnamefont
  {Van~Waeyenberge}}, \bibinfo {author} {\bibfnamefont {A.}~\bibnamefont
  {Vansteenkiste}}, \bibinfo {author} {\bibfnamefont {M.}~\bibnamefont
  {Weigand}}, \bibinfo {author} {\bibfnamefont {V.}~\bibnamefont {Sackmann}},
  \bibinfo {author} {\bibfnamefont {H.}~\bibnamefont {Stoll}}, \bibinfo
  {author} {\bibfnamefont {M.}~\bibnamefont {F\"ahnle}}, \bibinfo {author}
  {\bibfnamefont {T.}~\bibnamefont {Tyliszczak}}, \bibinfo {author}
  {\bibfnamefont {G.}~\bibnamefont {Woltersdorf}}, \bibinfo {author}
  {\bibfnamefont {C.~H.}\ \bibnamefont {Back}},\ and\ \bibinfo {author}
  {\bibfnamefont {G.}~\bibnamefont {Sch\"utz}},\ }\bibfield  {title} {\bibinfo
  {title} {Polarization selective magnetic vortex dynamics and core reversal in
  rotating magnetic fields},\ }\href@noop {} {\bibfield  {journal} {\bibinfo
  {journal} {Phys. Rev. Lett.}\ }\textbf {\bibinfo {volume} {101}},\ \bibinfo
  {pages} {197204} (\bibinfo {year} {2008})}\BibitemShut {NoStop}%
\bibitem [{\citenamefont {Liu}\ \emph {et~al.}(2007)\citenamefont {Liu},
  \citenamefont {Gliga}, \citenamefont {Hertel},\ and\ \citenamefont
  {Schneider}}]{liu2007current}%
  \BibitemOpen
  \bibfield  {author} {\bibinfo {author} {\bibfnamefont {Y.}~\bibnamefont
  {Liu}}, \bibinfo {author} {\bibfnamefont {S.}~\bibnamefont {Gliga}}, \bibinfo
  {author} {\bibfnamefont {R.}~\bibnamefont {Hertel}},\ and\ \bibinfo {author}
  {\bibfnamefont {C.}~\bibnamefont {Schneider}},\ }\bibfield  {title} {\bibinfo
  {title} {Current-induced magnetic vortex core switching in a permalloy
  nanodisk},\ }\href@noop {} {\bibfield  {journal} {\bibinfo  {journal} {Appl.
  Phys. Lett.}\ }\textbf {\bibinfo {volume} {91}} (\bibinfo {year}
  {2007})}\BibitemShut {NoStop}%
\bibitem [{\citenamefont {Fu}\ \emph {et~al.}(2018)\citenamefont {Fu},
  \citenamefont {Pollard}, \citenamefont {Chen}, \citenamefont {Yoo},
  \citenamefont {Yang},\ and\ \citenamefont {Zhu}}]{fu2018optical}%
  \BibitemOpen
  \bibfield  {author} {\bibinfo {author} {\bibfnamefont {X.}~\bibnamefont
  {Fu}}, \bibinfo {author} {\bibfnamefont {S.~D.}\ \bibnamefont {Pollard}},
  \bibinfo {author} {\bibfnamefont {B.}~\bibnamefont {Chen}}, \bibinfo {author}
  {\bibfnamefont {B.-K.}\ \bibnamefont {Yoo}}, \bibinfo {author} {\bibfnamefont
  {H.}~\bibnamefont {Yang}},\ and\ \bibinfo {author} {\bibfnamefont
  {Y.}~\bibnamefont {Zhu}},\ }\bibfield  {title} {\bibinfo {title} {Optical
  manipulation of magnetic vortices visualized in situ by lorentz electron
  microscopy},\ }\href@noop {} {\bibfield  {journal} {\bibinfo  {journal} {Sci.
  Adv.}\ }\textbf {\bibinfo {volume} {4}},\ \bibinfo {pages} {eaat3077}
  (\bibinfo {year} {2018})}\BibitemShut {NoStop}%
\bibitem [{\citenamefont {Yamada}\ \emph {et~al.}(2010)\citenamefont {Yamada},
  \citenamefont {Kasai}, \citenamefont {Nakatani}, \citenamefont {Kobayashi},\
  and\ \citenamefont {Ono}}]{yamada2010current}%
  \BibitemOpen
  \bibfield  {author} {\bibinfo {author} {\bibfnamefont {K.}~\bibnamefont
  {Yamada}}, \bibinfo {author} {\bibfnamefont {S.}~\bibnamefont {Kasai}},
  \bibinfo {author} {\bibfnamefont {Y.}~\bibnamefont {Nakatani}}, \bibinfo
  {author} {\bibfnamefont {K.}~\bibnamefont {Kobayashi}},\ and\ \bibinfo
  {author} {\bibfnamefont {T.}~\bibnamefont {Ono}},\ }\bibfield  {title}
  {\bibinfo {title} {Current-induced switching of magnetic vortex core in
  ferromagnetic elliptical disks},\ }\href@noop {} {\bibfield  {journal}
  {\bibinfo  {journal} {Appl. Phys. Lett.}\ }\textbf {\bibinfo {volume} {96}}
  (\bibinfo {year} {2010})}\BibitemShut {NoStop}%
\bibitem [{\citenamefont {Kravchuk}\ \emph {et~al.}(2009)\citenamefont
  {Kravchuk}, \citenamefont {Gaididei},\ and\ \citenamefont
  {Sheka}}]{PhysRevB.80.100405}%
  \BibitemOpen
  \bibfield  {author} {\bibinfo {author} {\bibfnamefont {V.~P.}\ \bibnamefont
  {Kravchuk}}, \bibinfo {author} {\bibfnamefont {Y.}~\bibnamefont {Gaididei}},\
  and\ \bibinfo {author} {\bibfnamefont {D.~D.}\ \bibnamefont {Sheka}},\
  }\bibfield  {title} {\bibinfo {title} {Nucleation of a vortex-antivortex pair
  in the presence of an immobile magnetic vortex},\ }\href@noop {} {\bibfield
  {journal} {\bibinfo  {journal} {Phys. Rev. B}\ }\textbf {\bibinfo {volume}
  {80}},\ \bibinfo {pages} {100405} (\bibinfo {year} {2009})}\BibitemShut
  {NoStop}%
\bibitem [{\citenamefont {Kammerer}\ \emph {et~al.}(2011)\citenamefont
  {Kammerer}, \citenamefont {Weigand}, \citenamefont {Curcic}, \citenamefont
  {Noske}, \citenamefont {Sproll}, \citenamefont {Vansteenkiste}, \citenamefont
  {Van~Waeyenberge}, \citenamefont {Stoll}, \citenamefont {Woltersdorf},
  \citenamefont {Back} \emph {et~al.}}]{kammerer2011magnetic}%
  \BibitemOpen
  \bibfield  {author} {\bibinfo {author} {\bibfnamefont {M.}~\bibnamefont
  {Kammerer}}, \bibinfo {author} {\bibfnamefont {M.}~\bibnamefont {Weigand}},
  \bibinfo {author} {\bibfnamefont {M.}~\bibnamefont {Curcic}}, \bibinfo
  {author} {\bibfnamefont {M.}~\bibnamefont {Noske}}, \bibinfo {author}
  {\bibfnamefont {M.}~\bibnamefont {Sproll}}, \bibinfo {author} {\bibfnamefont
  {A.}~\bibnamefont {Vansteenkiste}}, \bibinfo {author} {\bibfnamefont
  {B.}~\bibnamefont {Van~Waeyenberge}}, \bibinfo {author} {\bibfnamefont
  {H.}~\bibnamefont {Stoll}}, \bibinfo {author} {\bibfnamefont
  {G.}~\bibnamefont {Woltersdorf}}, \bibinfo {author} {\bibfnamefont {C.~H.}\
  \bibnamefont {Back}}, \emph {et~al.},\ }\bibfield  {title} {\bibinfo {title}
  {Magnetic vortex core reversal by excitation of spin waves},\ }\href@noop {}
  {\bibfield  {journal} {\bibinfo  {journal} {Nat. Commun.}\ }\textbf {\bibinfo
  {volume} {2}},\ \bibinfo {pages} {279} (\bibinfo {year} {2011})}\BibitemShut
  {NoStop}%
\bibitem [{\citenamefont {Hertel}\ \emph {et~al.}(2007)\citenamefont {Hertel},
  \citenamefont {Gliga}, \citenamefont {F\"ahnle},\ and\ \citenamefont
  {Schneider}}]{PhysRevLett.98.117201}%
  \BibitemOpen
  \bibfield  {author} {\bibinfo {author} {\bibfnamefont {R.}~\bibnamefont
  {Hertel}}, \bibinfo {author} {\bibfnamefont {S.}~\bibnamefont {Gliga}},
  \bibinfo {author} {\bibfnamefont {M.}~\bibnamefont {F\"ahnle}},\ and\
  \bibinfo {author} {\bibfnamefont {C.~M.}\ \bibnamefont {Schneider}},\
  }\bibfield  {title} {\bibinfo {title} {Ultrafast nanomagnetic toggle
  switching of vortex cores},\ }\href@noop {} {\bibfield  {journal} {\bibinfo
  {journal} {Phys. Rev. Lett.}\ }\textbf {\bibinfo {volume} {98}},\ \bibinfo
  {pages} {117201} (\bibinfo {year} {2007})}\BibitemShut {NoStop}%
\bibitem [{\citenamefont {Kim}\ and\ \citenamefont
  {Tserkovnyak}(2017{\natexlab{a}})}]{kim2017fast}%
  \BibitemOpen
  \bibfield  {author} {\bibinfo {author} {\bibfnamefont {S.~K.}\ \bibnamefont
  {Kim}}\ and\ \bibinfo {author} {\bibfnamefont {Y.}~\bibnamefont
  {Tserkovnyak}},\ }\bibfield  {title} {\bibinfo {title} {Fast vortex
  oscillations in a ferrimagnetic disk near the angular momentum compensation
  point},\ }\href@noop {} {\bibfield  {journal} {\bibinfo  {journal} {Appl.
  Phys. Lett.}\ }\textbf {\bibinfo {volume} {111}} (\bibinfo {year}
  {2017}{\natexlab{a}})}\BibitemShut {NoStop}%
\bibitem [{\citenamefont {Shibata}\ \emph {et~al.}(2006)\citenamefont
  {Shibata}, \citenamefont {Nakatani}, \citenamefont {Tatara}, \citenamefont
  {Kohno},\ and\ \citenamefont {Otani}}]{shibata2006current}%
  \BibitemOpen
  \bibfield  {author} {\bibinfo {author} {\bibfnamefont {J.}~\bibnamefont
  {Shibata}}, \bibinfo {author} {\bibfnamefont {Y.}~\bibnamefont {Nakatani}},
  \bibinfo {author} {\bibfnamefont {G.}~\bibnamefont {Tatara}}, \bibinfo
  {author} {\bibfnamefont {H.}~\bibnamefont {Kohno}},\ and\ \bibinfo {author}
  {\bibfnamefont {Y.}~\bibnamefont {Otani}},\ }\bibfield  {title} {\bibinfo
  {title} {Current-induced magnetic vortex motion by spin-transfer torque},\
  }\href@noop {} {\bibfield  {journal} {\bibinfo  {journal} {Phys. Rev. B}\
  }\textbf {\bibinfo {volume} {73}},\ \bibinfo {pages} {020403} (\bibinfo
  {year} {2006})}\BibitemShut {NoStop}%
\bibitem [{\citenamefont {Gao}\ \emph {et~al.}(2023)\citenamefont {Gao},
  \citenamefont {Wang}, \citenamefont {Zhao}, \citenamefont {Wang},
  \citenamefont {Hu},\ and\ \citenamefont {Yan}}]{gao2023interplay}%
  \BibitemOpen
  \bibfield  {author} {\bibinfo {author} {\bibfnamefont {Z.}~\bibnamefont
  {Gao}}, \bibinfo {author} {\bibfnamefont {F.}~\bibnamefont {Wang}}, \bibinfo
  {author} {\bibfnamefont {X.}~\bibnamefont {Zhao}}, \bibinfo {author}
  {\bibfnamefont {T.}~\bibnamefont {Wang}}, \bibinfo {author} {\bibfnamefont
  {J.}~\bibnamefont {Hu}},\ and\ \bibinfo {author} {\bibfnamefont
  {P.}~\bibnamefont {Yan}},\ }\bibfield  {title} {\bibinfo {title} {Interplay
  between spin wave and magnetic vortex},\ }\href@noop {} {\bibfield  {journal}
  {\bibinfo  {journal} {Physical Review B}\ }\textbf {\bibinfo {volume}
  {107}},\ \bibinfo {pages} {214418} (\bibinfo {year} {2023})}\BibitemShut
  {NoStop}%
\bibitem [{\citenamefont {Li}\ \emph {et~al.}(2021{\natexlab{a}})\citenamefont
  {Li}, \citenamefont {Wang}, \citenamefont {Zhang}, \citenamefont {Cao},\ and\
  \citenamefont {Yan}}]{li2021third}%
  \BibitemOpen
  \bibfield  {author} {\bibinfo {author} {\bibfnamefont {Z.-X.}\ \bibnamefont
  {Li}}, \bibinfo {author} {\bibfnamefont {Z.}~\bibnamefont {Wang}}, \bibinfo
  {author} {\bibfnamefont {Z.}~\bibnamefont {Zhang}}, \bibinfo {author}
  {\bibfnamefont {Y.}~\bibnamefont {Cao}},\ and\ \bibinfo {author}
  {\bibfnamefont {P.}~\bibnamefont {Yan}},\ }\bibfield  {title} {\bibinfo
  {title} {Third-order topological insulator in three-dimensional lattice of
  magnetic vortices},\ }\href@noop {} {\bibfield  {journal} {\bibinfo
  {journal} {Phys. Rev. B}\ }\textbf {\bibinfo {volume} {103}},\ \bibinfo
  {pages} {214442} (\bibinfo {year} {2021}{\natexlab{a}})}\BibitemShut
  {NoStop}%
\bibitem [{\citenamefont {Pribiag}\ \emph {et~al.}(2007)\citenamefont
  {Pribiag}, \citenamefont {Krivorotov}, \citenamefont {Fuchs}, \citenamefont
  {Braganca}, \citenamefont {Ozatay}, \citenamefont {Sankey}, \citenamefont
  {Ralph},\ and\ \citenamefont {Buhrman}}]{pribiag2007magnetic}%
  \BibitemOpen
  \bibfield  {author} {\bibinfo {author} {\bibfnamefont {V.}~\bibnamefont
  {Pribiag}}, \bibinfo {author} {\bibfnamefont {I.}~\bibnamefont {Krivorotov}},
  \bibinfo {author} {\bibfnamefont {G.}~\bibnamefont {Fuchs}}, \bibinfo
  {author} {\bibfnamefont {P.}~\bibnamefont {Braganca}}, \bibinfo {author}
  {\bibfnamefont {O.}~\bibnamefont {Ozatay}}, \bibinfo {author} {\bibfnamefont
  {J.}~\bibnamefont {Sankey}}, \bibinfo {author} {\bibfnamefont
  {D.}~\bibnamefont {Ralph}},\ and\ \bibinfo {author} {\bibfnamefont
  {R.}~\bibnamefont {Buhrman}},\ }\bibfield  {title} {\bibinfo {title}
  {Magnetic vortex oscillator driven by dc spin-polarized current},\
  }\href@noop {} {\bibfield  {journal} {\bibinfo  {journal} {Nat. Phys.}\
  }\textbf {\bibinfo {volume} {3}},\ \bibinfo {pages} {498} (\bibinfo {year}
  {2007})}\BibitemShut {NoStop}%
\bibitem [{\citenamefont {Yu}\ \emph {et~al.}(2013)\citenamefont {Yu},
  \citenamefont {Han}, \citenamefont {Yoo}, \citenamefont {Lee}, \citenamefont
  {Choi}, \citenamefont {Jung}, \citenamefont {Lee}, \citenamefont {Im},
  \citenamefont {Fischer},\ and\ \citenamefont {Kim}}]{yu2013resonant}%
  \BibitemOpen
  \bibfield  {author} {\bibinfo {author} {\bibfnamefont {Y.-S.}\ \bibnamefont
  {Yu}}, \bibinfo {author} {\bibfnamefont {D.-S.}\ \bibnamefont {Han}},
  \bibinfo {author} {\bibfnamefont {M.-W.}\ \bibnamefont {Yoo}}, \bibinfo
  {author} {\bibfnamefont {K.-S.}\ \bibnamefont {Lee}}, \bibinfo {author}
  {\bibfnamefont {Y.-S.}\ \bibnamefont {Choi}}, \bibinfo {author}
  {\bibfnamefont {H.}~\bibnamefont {Jung}}, \bibinfo {author} {\bibfnamefont
  {J.}~\bibnamefont {Lee}}, \bibinfo {author} {\bibfnamefont {M.-Y.}\
  \bibnamefont {Im}}, \bibinfo {author} {\bibfnamefont {P.}~\bibnamefont
  {Fischer}},\ and\ \bibinfo {author} {\bibfnamefont {S.-K.}\ \bibnamefont
  {Kim}},\ }\bibfield  {title} {\bibinfo {title} {Resonant amplification of
  vortex-core oscillations by coherent magnetic-field pulses},\ }\href@noop {}
  {\bibfield  {journal} {\bibinfo  {journal} {Sci. Rep.}\ }\textbf {\bibinfo
  {volume} {3}},\ \bibinfo {pages} {1301} (\bibinfo {year} {2013})}\BibitemShut
  {NoStop}%
\bibitem [{\citenamefont {Vogel}\ \emph {et~al.}(2010)\citenamefont {Vogel},
  \citenamefont {Drews}, \citenamefont {Kamionka}, \citenamefont {Bolte},\ and\
  \citenamefont {Meier}}]{PhysRevLett.105.037201}%
  \BibitemOpen
  \bibfield  {author} {\bibinfo {author} {\bibfnamefont {A.}~\bibnamefont
  {Vogel}}, \bibinfo {author} {\bibfnamefont {A.}~\bibnamefont {Drews}},
  \bibinfo {author} {\bibfnamefont {T.}~\bibnamefont {Kamionka}}, \bibinfo
  {author} {\bibfnamefont {M.}~\bibnamefont {Bolte}},\ and\ \bibinfo {author}
  {\bibfnamefont {G.}~\bibnamefont {Meier}},\ }\bibfield  {title} {\bibinfo
  {title} {Influence of dipolar interaction on vortex dynamics in arrays of
  ferromagnetic disks},\ }\href@noop {} {\bibfield  {journal} {\bibinfo
  {journal} {Phys. Rev. Lett.}\ }\textbf {\bibinfo {volume} {105}},\ \bibinfo
  {pages} {037201} (\bibinfo {year} {2010})}\BibitemShut {NoStop}%
\bibitem [{\citenamefont {Sugimoto}\ \emph {et~al.}(2011)\citenamefont
  {Sugimoto}, \citenamefont {Fukuma}, \citenamefont {Kasai}, \citenamefont
  {Kimura}, \citenamefont {Barman},\ and\ \citenamefont
  {Otani}}]{PhysRevLett.106.197203}%
  \BibitemOpen
  \bibfield  {author} {\bibinfo {author} {\bibfnamefont {S.}~\bibnamefont
  {Sugimoto}}, \bibinfo {author} {\bibfnamefont {Y.}~\bibnamefont {Fukuma}},
  \bibinfo {author} {\bibfnamefont {S.}~\bibnamefont {Kasai}}, \bibinfo
  {author} {\bibfnamefont {T.}~\bibnamefont {Kimura}}, \bibinfo {author}
  {\bibfnamefont {A.}~\bibnamefont {Barman}},\ and\ \bibinfo {author}
  {\bibfnamefont {Y.}~\bibnamefont {Otani}},\ }\bibfield  {title} {\bibinfo
  {title} {Dynamics of coupled vortices in a pair of ferromagnetic disks},\
  }\href@noop {} {\bibfield  {journal} {\bibinfo  {journal} {Phys. Rev. Lett.}\
  }\textbf {\bibinfo {volume} {106}},\ \bibinfo {pages} {197203} (\bibinfo
  {year} {2011})}\BibitemShut {NoStop}%
\bibitem [{\citenamefont {Kuepferling}\ \emph {et~al.}(2023)\citenamefont
  {Kuepferling}, \citenamefont {Casiraghi}, \citenamefont {Soares},
  \citenamefont {Durin}, \citenamefont {Garcia-Sanchez}, \citenamefont {Chen},
  \citenamefont {Back}, \citenamefont {Marrows}, \citenamefont {Tacchi},\ and\
  \citenamefont {Carlotti}}]{RevModPhys.95.015003}%
  \BibitemOpen
  \bibfield  {author} {\bibinfo {author} {\bibfnamefont {M.}~\bibnamefont
  {Kuepferling}}, \bibinfo {author} {\bibfnamefont {A.}~\bibnamefont
  {Casiraghi}}, \bibinfo {author} {\bibfnamefont {G.}~\bibnamefont {Soares}},
  \bibinfo {author} {\bibfnamefont {G.}~\bibnamefont {Durin}}, \bibinfo
  {author} {\bibfnamefont {F.}~\bibnamefont {Garcia-Sanchez}}, \bibinfo
  {author} {\bibfnamefont {L.}~\bibnamefont {Chen}}, \bibinfo {author}
  {\bibfnamefont {C.~H.}\ \bibnamefont {Back}}, \bibinfo {author}
  {\bibfnamefont {C.~H.}\ \bibnamefont {Marrows}}, \bibinfo {author}
  {\bibfnamefont {S.}~\bibnamefont {Tacchi}},\ and\ \bibinfo {author}
  {\bibfnamefont {G.}~\bibnamefont {Carlotti}},\ }\bibfield  {title} {\bibinfo
  {title} {Measuring interfacial dzyaloshinskii-moriya interaction in ultrathin
  magnetic films},\ }\href@noop {} {\bibfield  {journal} {\bibinfo  {journal}
  {Rev. Mod. Phys.}\ }\textbf {\bibinfo {volume} {95}},\ \bibinfo {pages}
  {015003} (\bibinfo {year} {2023})}\BibitemShut {NoStop}%
\bibitem [{\citenamefont {Tserkovnyak}\ and\ \citenamefont
  {Zou}(2019)}]{quantumvortex}%
  \BibitemOpen
  \bibfield  {author} {\bibinfo {author} {\bibfnamefont {Y.}~\bibnamefont
  {Tserkovnyak}}\ and\ \bibinfo {author} {\bibfnamefont {J.}~\bibnamefont
  {Zou}},\ }\bibfield  {title} {\bibinfo {title} {Quantum hydrodynamics of
  vorticity},\ }\href@noop {} {\bibfield  {journal} {\bibinfo  {journal} {Phys.
  Rev. Research}\ }\textbf {\bibinfo {volume} {1}},\ \bibinfo {pages} {033071}
  (\bibinfo {year} {2019})}\BibitemShut {NoStop}%
\bibitem [{\citenamefont {Akanda}\ \emph {et~al.}(2020)\citenamefont {Akanda},
  \citenamefont {Park},\ and\ \citenamefont {Lake}}]{PhysRevB.102.224414}%
  \BibitemOpen
  \bibfield  {author} {\bibinfo {author} {\bibfnamefont {M.~R.~K.}\
  \bibnamefont {Akanda}}, \bibinfo {author} {\bibfnamefont {I.~J.}\
  \bibnamefont {Park}},\ and\ \bibinfo {author} {\bibfnamefont {R.~K.}\
  \bibnamefont {Lake}},\ }\bibfield  {title} {\bibinfo {title} {Interfacial
  dzyaloshinskii-moriya interaction of antiferromagnetic materials},\
  }\href@noop {} {\bibfield  {journal} {\bibinfo  {journal} {Phys. Rev. B}\
  }\textbf {\bibinfo {volume} {102}},\ \bibinfo {pages} {224414} (\bibinfo
  {year} {2020})}\BibitemShut {NoStop}%
\bibitem [{\citenamefont {Qaiumzadeh}\ \emph {et~al.}(2018)\citenamefont
  {Qaiumzadeh}, \citenamefont {Ado}, \citenamefont {Duine}, \citenamefont
  {Titov},\ and\ \citenamefont {Brataas}}]{PhysRevLett.120.197202}%
  \BibitemOpen
  \bibfield  {author} {\bibinfo {author} {\bibfnamefont {A.}~\bibnamefont
  {Qaiumzadeh}}, \bibinfo {author} {\bibfnamefont {I.~A.}\ \bibnamefont {Ado}},
  \bibinfo {author} {\bibfnamefont {R.~A.}\ \bibnamefont {Duine}}, \bibinfo
  {author} {\bibfnamefont {M.}~\bibnamefont {Titov}},\ and\ \bibinfo {author}
  {\bibfnamefont {A.}~\bibnamefont {Brataas}},\ }\bibfield  {title} {\bibinfo
  {title} {Theory of the interfacial dzyaloshinskii-moriya interaction in
  rashba antiferromagnets},\ }\href@noop {} {\bibfield  {journal} {\bibinfo
  {journal} {Phys. Rev. Lett.}\ }\textbf {\bibinfo {volume} {120}},\ \bibinfo
  {pages} {197202} (\bibinfo {year} {2018})}\BibitemShut {NoStop}%
\bibitem [{\citenamefont {Volkov}\ \emph {et~al.}(2018)\citenamefont {Volkov},
  \citenamefont {Sheka}, \citenamefont {Gaididei}, \citenamefont {Kravchuk},
  \citenamefont {R{\"o}{\ss}ler}, \citenamefont {Fassbender},\ and\
  \citenamefont {Makarov}}]{volkov2018mesoscale}%
  \BibitemOpen
  \bibfield  {author} {\bibinfo {author} {\bibfnamefont {O.~M.}\ \bibnamefont
  {Volkov}}, \bibinfo {author} {\bibfnamefont {D.~D.}\ \bibnamefont {Sheka}},
  \bibinfo {author} {\bibfnamefont {Y.}~\bibnamefont {Gaididei}}, \bibinfo
  {author} {\bibfnamefont {V.~P.}\ \bibnamefont {Kravchuk}}, \bibinfo {author}
  {\bibfnamefont {U.~K.}\ \bibnamefont {R{\"o}{\ss}ler}}, \bibinfo {author}
  {\bibfnamefont {J.}~\bibnamefont {Fassbender}},\ and\ \bibinfo {author}
  {\bibfnamefont {D.}~\bibnamefont {Makarov}},\ }\bibfield  {title} {\bibinfo
  {title} {Mesoscale dzyaloshinskii-moriya interaction: geometrical tailoring
  of the magnetochirality},\ }\href@noop {} {\bibfield  {journal} {\bibinfo
  {journal} {Sci. Rep.}\ }\textbf {\bibinfo {volume} {8}},\ \bibinfo {pages}
  {866} (\bibinfo {year} {2018})}\BibitemShut {NoStop}%
\bibitem [{\citenamefont {Pethick}\ and\ \citenamefont
  {Smith}(2008)}]{pethick2008bose}%
  \BibitemOpen
  \bibfield  {author} {\bibinfo {author} {\bibfnamefont {C.~J.}\ \bibnamefont
  {Pethick}}\ and\ \bibinfo {author} {\bibfnamefont {H.}~\bibnamefont
  {Smith}},\ }\href@noop {} {\emph {\bibinfo {title} {Bose--Einstein
  condensation in dilute gases}}}\ (\bibinfo  {publisher} {Cambridge university
  press},\ \bibinfo {year} {2008})\BibitemShut {NoStop}%
\bibitem [{afm({\natexlab{a}})}]{afm_radius}%
  \BibitemOpen
  \href@noop {} {} \ \bibinfo {note} {One can also consider a
  disk with a larger radius. In this case, the equilibrium position of the
  vortex may not be at the center of the disk.}\BibitemShut {Stop}%
\bibitem [{\citenamefont {Guslienko}\ \emph {et~al.}(2001)\citenamefont
  {Guslienko}, \citenamefont {Novosad}, \citenamefont {Otani}, \citenamefont
  {Shima},\ and\ \citenamefont {Fukamichi}}]{guslienko2001field}%
  \BibitemOpen
  \bibfield  {author} {\bibinfo {author} {\bibfnamefont {K.~Y.}\ \bibnamefont
  {Guslienko}}, \bibinfo {author} {\bibfnamefont {V.}~\bibnamefont {Novosad}},
  \bibinfo {author} {\bibfnamefont {Y.}~\bibnamefont {Otani}}, \bibinfo
  {author} {\bibfnamefont {H.}~\bibnamefont {Shima}},\ and\ \bibinfo {author}
  {\bibfnamefont {K.}~\bibnamefont {Fukamichi}},\ }\bibfield  {title} {\bibinfo
  {title} {Field evolution of magnetic vortex state in ferromagnetic disks},\
  }\href@noop {} {\bibfield  {journal} {\bibinfo  {journal} {Appl. Phys.
  Lett.}\ }\textbf {\bibinfo {volume} {78}},\ \bibinfo {pages} {3848} (\bibinfo
  {year} {2001})}\BibitemShut {NoStop}%
\bibitem [{afm({\natexlab{b}})}]{afm_value}%
  \BibitemOpen
  \href@noop {} {} \ \bibinfo {note} {Here, we have omitted
  the constant term independent of the vortex number. Specifically, for
  $\mathcal{N}=1$, the potential energy evaluates to be $\mathcal{U}=\pi
  (7/2-2\ln 2) \mathcal{J}-2\pi\mathcal{D}$.}\BibitemShut {Stop}%
\bibitem [{\citenamefont {Tveten}\ \emph {et~al.}(2014)\citenamefont {Tveten},
  \citenamefont {Qaiumzadeh},\ and\ \citenamefont
  {Brataas}}]{PhysRevLett.112.147204}%
  \BibitemOpen
  \bibfield  {author} {\bibinfo {author} {\bibfnamefont {E.~G.}\ \bibnamefont
  {Tveten}}, \bibinfo {author} {\bibfnamefont {A.}~\bibnamefont {Qaiumzadeh}},\
  and\ \bibinfo {author} {\bibfnamefont {A.}~\bibnamefont {Brataas}},\
  }\bibfield  {title} {\bibinfo {title} {Antiferromagnetic domain wall motion
  induced by spin waves},\ }\href@noop {} {\bibfield  {journal} {\bibinfo
  {journal} {Phys. Rev. Lett.}\ }\textbf {\bibinfo {volume} {112}},\ \bibinfo
  {pages} {147204} (\bibinfo {year} {2014})}\BibitemShut {NoStop}%
\bibitem [{\citenamefont {Machado}\ \emph {et~al.}(2017)\citenamefont
  {Machado}, \citenamefont {Ribeiro}, \citenamefont {Holanda}, \citenamefont
  {Rodr\'{\i}guez-Su\'arez}, \citenamefont {Azevedo},\ and\ \citenamefont
  {Rezende}}]{PhysRevB.95.104418}%
  \BibitemOpen
  \bibfield  {author} {\bibinfo {author} {\bibfnamefont {F.~L.~A.}\
  \bibnamefont {Machado}}, \bibinfo {author} {\bibfnamefont {P.~R.~T.}\
  \bibnamefont {Ribeiro}}, \bibinfo {author} {\bibfnamefont {J.}~\bibnamefont
  {Holanda}}, \bibinfo {author} {\bibfnamefont {R.~L.}\ \bibnamefont
  {Rodr\'{\i}guez-Su\'arez}}, \bibinfo {author} {\bibfnamefont
  {A.}~\bibnamefont {Azevedo}},\ and\ \bibinfo {author} {\bibfnamefont {S.~M.}\
  \bibnamefont {Rezende}},\ }\bibfield  {title} {\bibinfo {title} {Spin-flop
  transition in the easy-plane antiferromagnet nickel oxide},\ }\href@noop {}
  {\bibfield  {journal} {\bibinfo  {journal} {Phys. Rev. B}\ }\textbf {\bibinfo
  {volume} {95}},\ \bibinfo {pages} {104418} (\bibinfo {year}
  {2017})}\BibitemShut {NoStop}%
\bibitem [{\citenamefont {Bader}\ and\ \citenamefont
  {Parkin}(2010)}]{bader2010spintronics}%
  \BibitemOpen
  \bibfield  {author} {\bibinfo {author} {\bibfnamefont {S.~D.}\ \bibnamefont
  {Bader}}\ and\ \bibinfo {author} {\bibfnamefont {S.}~\bibnamefont {Parkin}},\
  }\bibfield  {title} {\bibinfo {title} {Spintronics},\ }\href@noop {}
  {\bibfield  {journal} {\bibinfo  {journal} {Annu. Rev. Condens. Matter
  Phys.}\ }\textbf {\bibinfo {volume} {1}},\ \bibinfo {pages} {71} (\bibinfo
  {year} {2010})}\BibitemShut {NoStop}%
\bibitem [{\citenamefont {Baltz}\ \emph {et~al.}(2018)\citenamefont {Baltz},
  \citenamefont {Manchon}, \citenamefont {Tsoi}, \citenamefont {Moriyama},
  \citenamefont {Ono},\ and\ \citenamefont
  {Tserkovnyak}}]{RevModPhys.90.015005}%
  \BibitemOpen
  \bibfield  {author} {\bibinfo {author} {\bibfnamefont {V.}~\bibnamefont
  {Baltz}}, \bibinfo {author} {\bibfnamefont {A.}~\bibnamefont {Manchon}},
  \bibinfo {author} {\bibfnamefont {M.}~\bibnamefont {Tsoi}}, \bibinfo {author}
  {\bibfnamefont {T.}~\bibnamefont {Moriyama}}, \bibinfo {author}
  {\bibfnamefont {T.}~\bibnamefont {Ono}},\ and\ \bibinfo {author}
  {\bibfnamefont {Y.}~\bibnamefont {Tserkovnyak}},\ }\bibfield  {title}
  {\bibinfo {title} {Antiferromagnetic spintronics},\ }\href@noop {} {\bibfield
   {journal} {\bibinfo  {journal} {Rev. Mod. Phys.}\ }\textbf {\bibinfo
  {volume} {90}},\ \bibinfo {pages} {015005} (\bibinfo {year}
  {2018})}\BibitemShut {NoStop}%
\bibitem [{\citenamefont {Cheng}\ \emph {et~al.}(2014)\citenamefont {Cheng},
  \citenamefont {Xiao}, \citenamefont {Niu},\ and\ \citenamefont
  {Brataas}}]{PhysRevLett.113.057601}%
  \BibitemOpen
  \bibfield  {author} {\bibinfo {author} {\bibfnamefont {R.}~\bibnamefont
  {Cheng}}, \bibinfo {author} {\bibfnamefont {J.}~\bibnamefont {Xiao}},
  \bibinfo {author} {\bibfnamefont {Q.}~\bibnamefont {Niu}},\ and\ \bibinfo
  {author} {\bibfnamefont {A.}~\bibnamefont {Brataas}},\ }\bibfield  {title}
  {\bibinfo {title} {Spin pumping and spin-transfer torques in
  antiferromagnets},\ }\href@noop {} {\bibfield  {journal} {\bibinfo  {journal}
  {Phys. Rev. Lett.}\ }\textbf {\bibinfo {volume} {113}},\ \bibinfo {pages}
  {057601} (\bibinfo {year} {2014})}\BibitemShut {NoStop}%
\bibitem [{\citenamefont {Hals}\ \emph {et~al.}(2011)\citenamefont {Hals},
  \citenamefont {Tserkovnyak},\ and\ \citenamefont
  {Brataas}}]{PhysRevLett.106.107206}%
  \BibitemOpen
  \bibfield  {author} {\bibinfo {author} {\bibfnamefont {K.~M.~D.}\
  \bibnamefont {Hals}}, \bibinfo {author} {\bibfnamefont {Y.}~\bibnamefont
  {Tserkovnyak}},\ and\ \bibinfo {author} {\bibfnamefont {A.}~\bibnamefont
  {Brataas}},\ }\bibfield  {title} {\bibinfo {title} {Phenomenology of
  current-induced dynamics in antiferromagnets},\ }\href@noop {} {\bibfield
  {journal} {\bibinfo  {journal} {Phys. Rev. Lett.}\ }\textbf {\bibinfo
  {volume} {106}},\ \bibinfo {pages} {107206} (\bibinfo {year}
  {2011})}\BibitemShut {NoStop}%
\bibitem [{\citenamefont {Swaving}\ and\ \citenamefont
  {Duine}(2011)}]{PhysRevB.83.054428}%
  \BibitemOpen
  \bibfield  {author} {\bibinfo {author} {\bibfnamefont {A.~C.}\ \bibnamefont
  {Swaving}}\ and\ \bibinfo {author} {\bibfnamefont {R.~A.}\ \bibnamefont
  {Duine}},\ }\bibfield  {title} {\bibinfo {title} {Current-induced torques in
  continuous antiferromagnetic textures},\ }\href@noop {} {\bibfield  {journal}
  {\bibinfo  {journal} {Phys. Rev. B}\ }\textbf {\bibinfo {volume} {83}},\
  \bibinfo {pages} {054428} (\bibinfo {year} {2011})}\BibitemShut {NoStop}%
\bibitem [{\citenamefont {Dasgupta}\ and\ \citenamefont
  {Zou}(2021)}]{ji_2021_neel}%
  \BibitemOpen
  \bibfield  {author} {\bibinfo {author} {\bibfnamefont {S.}~\bibnamefont
  {Dasgupta}}\ and\ \bibinfo {author} {\bibfnamefont {J.}~\bibnamefont {Zou}},\
  }\bibfield  {title} {\bibinfo {title} {Zeeman term for the n\'eel vector in a
  two sublattice antiferromagnet},\ }\href@noop {} {\bibfield  {journal}
  {\bibinfo  {journal} {Phys. Rev. B}\ }\textbf {\bibinfo {volume} {104}},\
  \bibinfo {pages} {064415} (\bibinfo {year} {2021})}\BibitemShut {NoStop}%
\bibitem [{\citenamefont {Dasgupta}\ \emph {et~al.}(2017)\citenamefont
  {Dasgupta}, \citenamefont {Kim},\ and\ \citenamefont
  {Tchernyshyov}}]{sayakgauge}%
  \BibitemOpen
  \bibfield  {author} {\bibinfo {author} {\bibfnamefont {S.}~\bibnamefont
  {Dasgupta}}, \bibinfo {author} {\bibfnamefont {S.~K.}\ \bibnamefont {Kim}},\
  and\ \bibinfo {author} {\bibfnamefont {O.}~\bibnamefont {Tchernyshyov}},\
  }\bibfield  {title} {\bibinfo {title} {Gauge fields and related forces in
  antiferromagnetic soliton physics},\ }\href@noop {} {\bibfield  {journal}
  {\bibinfo  {journal} {Phys. Rev. B}\ }\textbf {\bibinfo {volume} {95}},\
  \bibinfo {pages} {220407} (\bibinfo {year} {2017})}\BibitemShut {NoStop}%
\bibitem [{\citenamefont {DE~GENNES}\ and\ \citenamefont
  {MATRICON}(1964)}]{RevModPhys.36.45}%
  \BibitemOpen
  \bibfield  {author} {\bibinfo {author} {\bibfnamefont {P.~G.}\ \bibnamefont
  {DE~GENNES}}\ and\ \bibinfo {author} {\bibfnamefont {J.}~\bibnamefont
  {MATRICON}},\ }\bibfield  {title} {\bibinfo {title} {Collective modes of
  vortex lines in superconductors of the second kind},\ }\href@noop {}
  {\bibfield  {journal} {\bibinfo  {journal} {Rev. Mod. Phys.}\ }\textbf
  {\bibinfo {volume} {36}},\ \bibinfo {pages} {45} (\bibinfo {year}
  {1964})}\BibitemShut {NoStop}%
\bibitem [{afm({\natexlab{c}})}]{afm_sm}%
  \BibitemOpen
  \href@noop {} {} \ \bibinfo {note} {See Supplemental
  Material for (i) Derivation of vortex-vortex interaction and (ii) Trajectories of the vortex.}\BibitemShut {Stop}%
\bibitem [{\citenamefont {Trifunovic}\ \emph {et~al.}(2013)\citenamefont
  {Trifunovic}, \citenamefont {Pedrocchi},\ and\ \citenamefont
  {Loss}}]{Daniel2013prx}%
  \BibitemOpen
  \bibfield  {author} {\bibinfo {author} {\bibfnamefont {L.}~\bibnamefont
  {Trifunovic}}, \bibinfo {author} {\bibfnamefont {F.~L.}\ \bibnamefont
  {Pedrocchi}},\ and\ \bibinfo {author} {\bibfnamefont {D.}~\bibnamefont
  {Loss}},\ }\bibfield  {title} {\bibinfo {title} {Long-distance entanglement
  of spin qubits via ferromagnet},\ }\href@noop {} {\bibfield  {journal}
  {\bibinfo  {journal} {Phys. Rev. X}\ }\textbf {\bibinfo {volume} {3}},\
  \bibinfo {pages} {041023} (\bibinfo {year} {2013})}\BibitemShut {NoStop}%
\bibitem [{\citenamefont {Het{\'e}nyi}\ \emph {et~al.}(2022)\citenamefont
  {Het{\'e}nyi}, \citenamefont {Mook}, \citenamefont {Klinovaja},\ and\
  \citenamefont {Loss}}]{hetenyi2022long}%
  \BibitemOpen
  \bibfield  {author} {\bibinfo {author} {\bibfnamefont {B.}~\bibnamefont
  {Het{\'e}nyi}}, \bibinfo {author} {\bibfnamefont {A.}~\bibnamefont {Mook}},
  \bibinfo {author} {\bibfnamefont {J.}~\bibnamefont {Klinovaja}},\ and\
  \bibinfo {author} {\bibfnamefont {D.}~\bibnamefont {Loss}},\ }\bibfield
  {title} {\bibinfo {title} {Long-distance coupling of spin qubits via
  topological magnons},\ }\href@noop {} {\bibfield  {journal} {\bibinfo
  {journal} {Phys. Rev. B}\ }\textbf {\bibinfo {volume} {106}},\ \bibinfo
  {pages} {235409} (\bibinfo {year} {2022})}\BibitemShut {NoStop}%
\bibitem [{\citenamefont {Zou}\ \emph {et~al.}(2024)\citenamefont {Zou},
  \citenamefont {Bosco}, \citenamefont {Thingstad}, \citenamefont {Klinovaja},\
  and\ \citenamefont {Loss}}]{zou2024prl}%
  \BibitemOpen
  \bibfield  {author} {\bibinfo {author} {\bibfnamefont {J.}~\bibnamefont
  {Zou}}, \bibinfo {author} {\bibfnamefont {S.}~\bibnamefont {Bosco}}, \bibinfo
  {author} {\bibfnamefont {E.}~\bibnamefont {Thingstad}}, \bibinfo {author}
  {\bibfnamefont {J.}~\bibnamefont {Klinovaja}},\ and\ \bibinfo {author}
  {\bibfnamefont {D.}~\bibnamefont {Loss}},\ }\bibfield  {title} {\bibinfo
  {title} {Dissipative spin-wave diode and nonreciprocal magnonic amplifier},\
  }\href@noop {} {\bibfield  {journal} {\bibinfo  {journal} {Phys. Rev. Lett.}\
  }\textbf {\bibinfo {volume} {132}},\ \bibinfo {pages} {036701} (\bibinfo
  {year} {2024})}\BibitemShut {NoStop}%
\bibitem [{\citenamefont {Zou}\ \emph {et~al.}(2023{\natexlab{b}})\citenamefont
  {Zou}, \citenamefont {Bosco},\ and\ \citenamefont {Loss}}]{zou2023spatially}%
  \BibitemOpen
  \bibfield  {author} {\bibinfo {author} {\bibfnamefont {J.}~\bibnamefont
  {Zou}}, \bibinfo {author} {\bibfnamefont {S.}~\bibnamefont {Bosco}},\ and\
  \bibinfo {author} {\bibfnamefont {D.}~\bibnamefont {Loss}},\ }\bibfield
  {title} {\bibinfo {title} {Spatially correlated classical and quantum noise
  in driven qubits: The good, the bad, and the ugly},\ }\href@noop {}
  {\bibfield  {journal} {\bibinfo  {journal} {arXiv:2308.03054}\ } (\bibinfo
  {year} {2023}{\natexlab{b}})}\BibitemShut {NoStop}%
\bibitem [{\citenamefont {Bravyi}\ \emph {et~al.}(2011)\citenamefont {Bravyi},
  \citenamefont {DiVincenzo},\ and\ \citenamefont
  {Loss}}]{bravyi2011schrieffer}%
  \BibitemOpen
  \bibfield  {author} {\bibinfo {author} {\bibfnamefont {S.}~\bibnamefont
  {Bravyi}}, \bibinfo {author} {\bibfnamefont {D.~P.}\ \bibnamefont
  {DiVincenzo}},\ and\ \bibinfo {author} {\bibfnamefont {D.}~\bibnamefont
  {Loss}},\ }\bibfield  {title} {\bibinfo {title} {Schrieffer--wolff
  transformation for quantum many-body systems},\ }\href@noop {} {\bibfield
  {journal} {\bibinfo  {journal} {Ann. Phys.}\ }\textbf {\bibinfo {volume}
  {326}},\ \bibinfo {pages} {2793} (\bibinfo {year} {2011})}\BibitemShut
  {NoStop}%
\bibitem [{\citenamefont {Li}\ \emph {et~al.}(2021{\natexlab{b}})\citenamefont
  {Li}, \citenamefont {Cao},\ and\ \citenamefont {Yan}}]{li2021topological}%
  \BibitemOpen
  \bibfield  {author} {\bibinfo {author} {\bibfnamefont {Z.-X.}\ \bibnamefont
  {Li}}, \bibinfo {author} {\bibfnamefont {Y.}~\bibnamefont {Cao}},\ and\
  \bibinfo {author} {\bibfnamefont {P.}~\bibnamefont {Yan}},\ }\bibfield
  {title} {\bibinfo {title} {Topological insulators and semimetals in classical
  magnetic systems},\ }\href@noop {} {\bibfield  {journal} {\bibinfo  {journal}
  {Phys. Rep.}\ }\textbf {\bibinfo {volume} {915}},\ \bibinfo {pages} {1}
  (\bibinfo {year} {2021}{\natexlab{b}})}\BibitemShut {NoStop}%
\bibitem [{\citenamefont {Kim}\ and\ \citenamefont
  {Tserkovnyak}(2017{\natexlab{b}})}]{PhysRevLett.119.077204}%
  \BibitemOpen
  \bibfield  {author} {\bibinfo {author} {\bibfnamefont {S.~K.}\ \bibnamefont
  {Kim}}\ and\ \bibinfo {author} {\bibfnamefont {Y.}~\bibnamefont
  {Tserkovnyak}},\ }\bibfield  {title} {\bibinfo {title} {Chiral edge mode in
  the coupled dynamics of magnetic solitons in a honeycomb lattice},\
  }\href@noop {} {\bibfield  {journal} {\bibinfo  {journal} {Phys. Rev. Lett.}\
  }\textbf {\bibinfo {volume} {119}},\ \bibinfo {pages} {077204} (\bibinfo
  {year} {2017}{\natexlab{b}})}\BibitemShut {NoStop}%
\bibitem [{\citenamefont {Li}\ \emph {et~al.}(2019)\citenamefont {Li},
  \citenamefont {Cao}, \citenamefont {Yan},\ and\ \citenamefont
  {Wang}}]{li2019higher}%
  \BibitemOpen
  \bibfield  {author} {\bibinfo {author} {\bibfnamefont {Z.}~\bibnamefont
  {Li}}, \bibinfo {author} {\bibfnamefont {Y.}~\bibnamefont {Cao}}, \bibinfo
  {author} {\bibfnamefont {P.}~\bibnamefont {Yan}},\ and\ \bibinfo {author}
  {\bibfnamefont {X.}~\bibnamefont {Wang}},\ }\bibfield  {title} {\bibinfo
  {title} {Higher-order topological solitonic insulators},\ }\href@noop {}
  {\bibfield  {journal} {\bibinfo  {journal} {npj Computational Materials}\
  }\textbf {\bibinfo {volume} {5}},\ \bibinfo {pages} {107} (\bibinfo {year}
  {2019})}\BibitemShut {NoStop}%
\bibitem [{\citenamefont {Li}\ \emph {et~al.}(2018)\citenamefont {Li},
  \citenamefont {Wang}, \citenamefont {Cao},\ and\ \citenamefont
  {Yan}}]{PhysRevB.98.180407}%
  \BibitemOpen
  \bibfield  {author} {\bibinfo {author} {\bibfnamefont {Z.-X.}\ \bibnamefont
  {Li}}, \bibinfo {author} {\bibfnamefont {C.}~\bibnamefont {Wang}}, \bibinfo
  {author} {\bibfnamefont {Y.}~\bibnamefont {Cao}},\ and\ \bibinfo {author}
  {\bibfnamefont {P.}~\bibnamefont {Yan}},\ }\bibfield  {title} {\bibinfo
  {title} {Edge states in a two-dimensional honeycomb lattice of massive
  magnetic skyrmions},\ }\href@noop {} {\bibfield  {journal} {\bibinfo
  {journal} {Phys. Rev. B}\ }\textbf {\bibinfo {volume} {98}},\ \bibinfo
  {pages} {180407} (\bibinfo {year} {2018})}\BibitemShut {NoStop}%
\bibitem [{\citenamefont {Li}\ \emph {et~al.}(2021{\natexlab{c}})\citenamefont
  {Li}, \citenamefont {Wang}, \citenamefont {Cao}, \citenamefont {Zhang},\ and\
  \citenamefont {Yan}}]{PhysRevB.103.054438}%
  \BibitemOpen
  \bibfield  {author} {\bibinfo {author} {\bibfnamefont {Z.-X.}\ \bibnamefont
  {Li}}, \bibinfo {author} {\bibfnamefont {Z.}~\bibnamefont {Wang}}, \bibinfo
  {author} {\bibfnamefont {Y.}~\bibnamefont {Cao}}, \bibinfo {author}
  {\bibfnamefont {H.~W.}\ \bibnamefont {Zhang}},\ and\ \bibinfo {author}
  {\bibfnamefont {P.}~\bibnamefont {Yan}},\ }\bibfield  {title} {\bibinfo
  {title} {Robust edge states in magnetic soliton racetrack},\ }\href@noop {}
  {\bibfield  {journal} {\bibinfo  {journal} {Phys. Rev. B}\ }\textbf {\bibinfo
  {volume} {103}},\ \bibinfo {pages} {054438} (\bibinfo {year}
  {2021}{\natexlab{c}})}\BibitemShut {NoStop}%
\bibitem [{\citenamefont {Yu}\ \emph {et~al.}(2023)\citenamefont {Yu},
  \citenamefont {Zou}, \citenamefont {Zeng}, \citenamefont {Rao},\ and\
  \citenamefont {Xia}}]{yu2023non}%
  \BibitemOpen
  \bibfield  {author} {\bibinfo {author} {\bibfnamefont {T.}~\bibnamefont
  {Yu}}, \bibinfo {author} {\bibfnamefont {J.}~\bibnamefont {Zou}}, \bibinfo
  {author} {\bibfnamefont {B.}~\bibnamefont {Zeng}}, \bibinfo {author}
  {\bibfnamefont {J.}~\bibnamefont {Rao}},\ and\ \bibinfo {author}
  {\bibfnamefont {K.}~\bibnamefont {Xia}},\ }\bibfield  {title} {\bibinfo
  {title} {Non-hermitian topological magnonics},\ }\href@noop {} {\bibfield
  {journal} {\bibinfo  {journal} {arXiv:2306.04348}\ } (\bibinfo {year}
  {2023})}\BibitemShut {NoStop}%
\bibitem [{\citenamefont {Hurst}\ and\ \citenamefont
  {Flebus}(2022)}]{hurst2022non}%
  \BibitemOpen
  \bibfield  {author} {\bibinfo {author} {\bibfnamefont {H.~M.}\ \bibnamefont
  {Hurst}}\ and\ \bibinfo {author} {\bibfnamefont {B.}~\bibnamefont {Flebus}},\
  }\bibfield  {title} {\bibinfo {title} {Non-hermitian physics in magnetic
  systems},\ }\href@noop {} {\bibfield  {journal} {\bibinfo  {journal} {J.
  Appl. Phys.}\ }\textbf {\bibinfo {volume} {132}},\ \bibinfo {pages} {220902}
  (\bibinfo {year} {2022})}\BibitemShut {NoStop}%
\bibitem [{\citenamefont {Yu}\ \emph {et~al.}(2019)\citenamefont {Yu},
  \citenamefont {Wang}, \citenamefont {Yuan}, \citenamefont {Xiao} \emph
  {et~al.}}]{yu2019prediction}%
  \BibitemOpen
  \bibfield  {author} {\bibinfo {author} {\bibfnamefont {W.}~\bibnamefont
  {Yu}}, \bibinfo {author} {\bibfnamefont {J.}~\bibnamefont {Wang}}, \bibinfo
  {author} {\bibfnamefont {H.}~\bibnamefont {Yuan}}, \bibinfo {author}
  {\bibfnamefont {J.}~\bibnamefont {Xiao}}, \emph {et~al.},\ }\bibfield
  {title} {\bibinfo {title} {Prediction of attractive level crossing via a
  dissipative mode},\ }\href@noop {} {\bibfield  {journal} {\bibinfo  {journal}
  {Phys. Rev. Lett.}\ }\textbf {\bibinfo {volume} {123}},\ \bibinfo {pages}
  {227201} (\bibinfo {year} {2019})}\BibitemShut {NoStop}%
\bibitem [{\citenamefont {Zou}\ \emph {et~al.}(2022)\citenamefont {Zou},
  \citenamefont {Zhang},\ and\ \citenamefont {Tserkovnyak}}]{zou2022prb}%
  \BibitemOpen
  \bibfield  {author} {\bibinfo {author} {\bibfnamefont {J.}~\bibnamefont
  {Zou}}, \bibinfo {author} {\bibfnamefont {S.}~\bibnamefont {Zhang}},\ and\
  \bibinfo {author} {\bibfnamefont {Y.}~\bibnamefont {Tserkovnyak}},\
  }\bibfield  {title} {\bibinfo {title} {Bell-state generation for spin qubits
  via dissipative coupling},\ }\href@noop {} {\bibfield  {journal} {\bibinfo
  {journal} {Phys. Rev. B}\ }\textbf {\bibinfo {volume} {106}},\ \bibinfo
  {pages} {L180406} (\bibinfo {year} {2022})}\BibitemShut {NoStop}%
\bibitem [{\citenamefont {Ashida}\ \emph {et~al.}(2020)\citenamefont {Ashida},
  \citenamefont {Gong},\ and\ \citenamefont {Ueda}}]{ashida2020non}%
  \BibitemOpen
  \bibfield  {author} {\bibinfo {author} {\bibfnamefont {Y.}~\bibnamefont
  {Ashida}}, \bibinfo {author} {\bibfnamefont {Z.}~\bibnamefont {Gong}},\ and\
  \bibinfo {author} {\bibfnamefont {M.}~\bibnamefont {Ueda}},\ }\bibfield
  {title} {\bibinfo {title} {Non-hermitian physics},\ }\href@noop {} {\bibfield
   {journal} {\bibinfo  {journal} {Adv. Phys.}\ }\textbf {\bibinfo {volume}
  {69}},\ \bibinfo {pages} {249} (\bibinfo {year} {2020})}\BibitemShut
  {NoStop}%
\bibitem [{\citenamefont {Nakata}\ \emph {et~al.}(2024)\citenamefont {Nakata},
  \citenamefont {Zou}, \citenamefont {Klinovaja},\ and\ \citenamefont
  {Loss}}]{nakata2024magnonic}%
  \BibitemOpen
  \bibfield  {author} {\bibinfo {author} {\bibfnamefont {K.}~\bibnamefont
  {Nakata}}, \bibinfo {author} {\bibfnamefont {J.}~\bibnamefont {Zou}},
  \bibinfo {author} {\bibfnamefont {J.}~\bibnamefont {Klinovaja}},\ and\
  \bibinfo {author} {\bibfnamefont {D.}~\bibnamefont {Loss}},\ }\bibfield
  {title} {\bibinfo {title} {Magnonic $\varphi$ josephson junctions and
  synchronized precession},\ }\href@noop {} {\bibfield  {journal} {\bibinfo
  {journal} {arXiv:2403.01625}\ } (\bibinfo {year} {2024})}\BibitemShut
  {NoStop}%
\bibitem [{\citenamefont {Yuan}\ \emph {et~al.}(2022)\citenamefont {Yuan},
  \citenamefont {Cao}, \citenamefont {Kamra}, \citenamefont {Duine},\ and\
  \citenamefont {Yan}}]{YUAN20221}%
  \BibitemOpen
  \bibfield  {author} {\bibinfo {author} {\bibfnamefont {H.}~\bibnamefont
  {Yuan}}, \bibinfo {author} {\bibfnamefont {Y.}~\bibnamefont {Cao}}, \bibinfo
  {author} {\bibfnamefont {A.}~\bibnamefont {Kamra}}, \bibinfo {author}
  {\bibfnamefont {R.~A.}\ \bibnamefont {Duine}},\ and\ \bibinfo {author}
  {\bibfnamefont {P.}~\bibnamefont {Yan}},\ }\bibfield  {title} {\bibinfo
  {title} {Quantum magnonics: When magnon spintronics meets quantum information
  science},\ }\href@noop {} {\bibfield  {journal} {\bibinfo  {journal} {Phys.
  Rep.}\ }\textbf {\bibinfo {volume} {965}},\ \bibinfo {pages} {1} (\bibinfo
  {year} {2022})}\BibitemShut {NoStop}%
\end{thebibliography}

\begin{thebibliography}{2}%
\makeatletter
\providecommand \@ifxundefined [1]{%
 \@ifx{#1\undefined}
}%
\providecommand \@ifnum [1]{%
 \ifnum #1\expandafter \@firstoftwo
 \else \expandafter \@secondoftwo
 \fi
}%
\providecommand \@ifx [1]{%
 \ifx #1\expandafter \@firstoftwo
 \else \expandafter \@secondoftwo
 \fi
}%
\providecommand \natexlab [1]{#1}%
\providecommand \enquote  [1]{``#1''}%
\providecommand \bibnamefont  [1]{#1}%
\providecommand \bibfnamefont [1]{#1}%
\providecommand \citenamefont [1]{#1}%
\providecommand \href@noop [0]{\@secondoftwo}%
\providecommand \href [0]{\begingroup \@sanitize@url \@href}%
\providecommand \@href[1]{\@@startlink{#1}\@@href}%
\providecommand \@@href[1]{\endgroup#1\@@endlink}%
\providecommand \@sanitize@url [0]{\catcode `\\12\catcode `\$12\catcode
  `\&12\catcode `\#12\catcode `\^12\catcode `\_12\catcode `\%12\relax}%
\providecommand \@@startlink[1]{}%
\providecommand \@@endlink[0]{}%
\providecommand \url  [0]{\begingroup\@sanitize@url \@url }%
\providecommand \@url [1]{\endgroup\@href {#1}{\urlprefix }}%
\providecommand \urlprefix  [0]{URL }%
\providecommand \Eprint [0]{\href }%
\providecommand \doibase [0]{https://doi.org/}%
\providecommand \selectlanguage [0]{\@gobble}%
\providecommand \bibinfo  [0]{\@secondoftwo}%
\providecommand \bibfield  [0]{\@secondoftwo}%
\providecommand \translation [1]{[#1]}%
\providecommand \BibitemOpen [0]{}%
\providecommand \bibitemStop [0]{}%
\providecommand \bibitemNoStop [0]{.\EOS\space}%
\providecommand \EOS [0]{\spacefactor3000\relax}%
\providecommand \BibitemShut  [1]{\csname bibitem#1\endcsname}%
\let\auto@bib@innerbib\@empty
\bibitem [{\citenamefont {Bravyi}\ \emph {et~al.}(2011)\citenamefont {Bravyi},
  \citenamefont {DiVincenzo},\ and\ \citenamefont
  {Loss}}]{bravyi2011schrieffer}%
  \BibitemOpen
  \bibfield  {author} {\bibinfo {author} {\bibfnamefont {S.}~\bibnamefont
  {Bravyi}}, \bibinfo {author} {\bibfnamefont {D.~P.}\ \bibnamefont
  {DiVincenzo}},\ and\ \bibinfo {author} {\bibfnamefont {D.}~\bibnamefont
  {Loss}},\ }\bibfield  {title} {\bibinfo {title} {Schrieffer--wolff
  transformation for quantum many-body systems},\ }\href@noop {} {\bibfield
  {journal} {\bibinfo  {journal} {Annals of physics}\ }\textbf {\bibinfo
  {volume} {326}},\ \bibinfo {pages} {2793} (\bibinfo {year}
  {2011})}\BibitemShut {NoStop}%
\bibitem [{\citenamefont {Trifunovic}\ \emph {et~al.}(2013)\citenamefont
  {Trifunovic}, \citenamefont {Pedrocchi},\ and\ \citenamefont
  {Loss}}]{Daniel2013prx}%
  \BibitemOpen
  \bibfield  {author} {\bibinfo {author} {\bibfnamefont {L.}~\bibnamefont
  {Trifunovic}}, \bibinfo {author} {\bibfnamefont {F.~L.}\ \bibnamefont
  {Pedrocchi}},\ and\ \bibinfo {author} {\bibfnamefont {D.}~\bibnamefont
  {Loss}},\ }\bibfield  {title} {\bibinfo {title} {Long-distance entanglement
  of spin qubits via ferromagnet},\ }\href@noop {} {\bibfield  {journal}
  {\bibinfo  {journal} {Phys. Rev. X}\ }\textbf {\bibinfo {volume} {3}},\
  \bibinfo {pages} {041023} (\bibinfo {year} {2013})}\BibitemShut {NoStop}%
\end{thebibliography}
%

\end{document}